\documentclass[%
 aip,
 amsmath,amssymb,
 reprint,%
]{revtex4-1}

\usepackage[scaled=.92]{helvet}

\usepackage{stix2}
\usepackage{microtype}

\usepackage{graphicx}
\usepackage{dcolumn}
\usepackage{bm}

\usepackage{makecell}

\newcommand{\curl}{\nabla \times}

\newcommand{\B}{{\bf{B}}}
\newcommand{\E}{{\bf E}}
\newcommand{\J}{{\bf J}}
\newcommand{\V}{{\bf V}}

\begin{document}

\title{Modeling Intense-Electron-Beam Generated Plasmas Using a Rigid-Beam Approximation}

\author{A.~S.~Richardson}
\email{steve.richardson@nrl.navy.mil}
\author{S.~B.~Swanekamp}%
\affiliation{ 
Plasma Physics Division, Naval Research Laboratory, Washington, DC 20375, USA
}%
\author{N.~D.~Isner}
\affiliation{%
Syntek Technologies, Fairfax, VA 22031, USA
}%

\author{D.~D.~Hinshelwood}
\affiliation{ 
Plasma Physics Division, Naval Research Laboratory, Washington, DC 20375, USA
}%

\author{D.~Mosher}
\affiliation{%
Syntek Technologies, Fairfax, VA 22031, USA
}%

\author{P.~E.~Adamson}
\author{I.~M.~Rittersdorf}
\author{Tz.~B.~Petrova}
\affiliation{ 
Plasma Physics Division, Naval Research Laboratory, Washington, DC 20375, USA
}%

\author{D.~J.~Watkins}
\affiliation{%
Syntek Technologies, Fairfax, VA 22031, USA
}%

\date{August 22, 2021}

\begin{abstract}
A model of an electron-beam-plasma system is introduced to model the electrical breakdown physics of low-pressure nitrogen irradiated by an intense pulsed electron beam. The rapidly rising beam current induces an electric field which drives a return current in the plasma. The {\em rigid-beam model} is a reduction of the problem geometry to cylindrical coordinates and simplifications to Maxwell’s equations that are driven by a prescribed electron beam current density. The model is convenient for comparing various reductions of the plasma dynamics and plasma chemistry while maintaining a good approximation to the overall magnitude of the beam-created electric field. The usefulness of this model is demonstrated by coupling the rigid-beam model to a fluid plasma model and a simplified nitrogen plasma chemistry. The dynamics of this coupled system are computed for a range of background gas pressures, and the results are compared with experimental measurements.
At pressures 1~Torr and above, the simulated line-integrated electron densities are within a factor of two of measurements, and show the same trend with pressure as observed in experiment.
\end{abstract}

\maketitle

\section{\label{sec:intro}Introduction}

The interaction of an intense electron beam (e-beam) with partially or fully ionized gas has many applications including plasma processing,\cite{walton04,Leonhardt_2007,Boris_2013} interactions in the ionosphere (e.g. solar wind and aurora),\cite{10.1029/165GM03,10.1029/2011SW000734} and electromagnetic pulse events.\cite{4328130,4091115,lazarev2005} In many of these natural and laboratory plasmas, air and its properties play an important role. The rapid change of the air from an insulator to an electrically conductive plasma in the presence of an intense e-beam is a complex phenomenon which depends on the plasma chemistry and rapidly changing electromagnetic fields. In this paper, a set of approximations to the beam and electromagnetic field equations, called the rigid-beam model, are introduced. This model provides a convenient framework for examining the plasma response as system parameters are varied or when different plasma approximations are made. 

At early times during an e-beam pulse, the beam’s space-charge creates an electrostatic field in the gas that can cause the beam to expand radially at the front. Beam-impact ionization then produces sufficient plasma density to effectively neutralize the beam's space charge so that it can be confined radially and propagate axially.\cite{10.1063/1.860088}  Once charge neutralization occurs, the electric field is dominated by an inductive component driven by the rapidly changing e-beam current.  Electrical breakdown in this field, together with beam-impact ionization, rapidly increases the plasma density, and a plasma return current begins to flow in the increasingly conductive plasma. This plasma return current can be comparable to the beam current and contributes to the duration and magnitude of the inductive electric field inside the plasma and the electromagnetic field radiating outside of the beam-plasma system.  The processes that cause the air to evolve from a low conductivity gas to a highly conductive plasma are quite complex, and involve e-beam dynamics, physical and chemical changes in the gas, thermal plasma dynamics, and collisional interactions between the plasma and the gas. The beam and plasma are coupled through Maxwell’s equations. To help better understand the plasma physics, it is useful to develop reduced models of the beam-plasma system that capture the essential physics. These reduced models offer relatively fast computation times, which allows us to rapidly develop plasma chemistry models and to determine pressures where a kinetic plasma model is needed and where a fluid model can suffice.

The rigid-beam (RB) model described here is a reduction of the problem geometry to cylindrical coordinates with the e-beam treated as a specified source that drives the remaining system of equations, i.e. beam dynamics are not coupled into the governing equations.  The model allows the e-beam to have arbitrary time variations of current and voltage and an arbitrary current density profile. In this paper, we use spatial and time variations guided by experiments. The electron beam used in experiments has a peak current of about 4~kA, a peak voltage of about 100~kV, a pulse width of about 50~ns, and a diameter of about 4~cm.\cite{ddh_beams}
 The model described in this paper is an extension of previous models developed to study the interaction of an intense e-beam with nitrogen gas.\cite{doi:10.1063/1.1694245,doi:10.1063/1.345772,doi:10.1063/1.4950840} The e-beam is assumed to be injected parallel to the axis of a perfectly conducting cylinder of radius $R_w$ filled with low-pressure nitrogen in the 1 to 10~Torr range. The usefulness of an earlier RB model was demonstrated by comparing two models for electron-gas collisions: one with collision rates computed from the reduced electric field ($E/p$) and a second where they were determined from the mean plasma electron energy.\cite{swanekamp2021rigidbeam} The plasma chemistry model used in this paper adds dissociative recombination and beam-impact ionization, both of which were neglected in that earlier RB model. {In addition to this expanded plasma chemistry, the model presented in this paper extends the 0-dimensional RB model of Ref.~\onlinecite{doi:10.1063/1.4950840} to a 1-dimensional model by allowing an arbitrary radial profile for the electron beam.}
The RB model employed in this paper is described more fully in Sec.~\ref{rb_model}. 

The complete set of collisions that affect the plasma and gas particles is referred to as the plasma chemistry. Here, the plasma chemistry will be limited to that of a weakly ionized plasma. “Weakly ionized” implies that the collision rate between electrons and molecular nitrogen in the ${\rm N}_2 ({\rm X}\,^1\Sigma)$ ground state remains much larger than all other collisional rates. 
The various cross sections and rates required to model weakly ionized N$_2$ discharges are readily available through tables in the literature,\cite{PhysRevA.31.2932}  the LXCat database,\cite{phelps_lxcat} and the Quantemol database QDB.\cite{Tennyson_2017} 
The chemistry model is described in more detail in Sec.~\ref{plasma_response}. Results from fluid simulations using the RB model over a range of gas pressures are presented in Sec.~\ref{sec_results}.

The RB model described in this paper extends the 0-D model of Ref.~\onlinecite{doi:10.1063/1.4950840} to a 1-D model, and expands the chemistry used in Ref.~\onlinecite{swanekamp2021rigidbeam} to include recombination and beam-impact ionization. These additions allow simulations to be performed which maintain a higher fidelity with the experiments.
The line-integrated electron densities predicted by the simulations and reported in Sec.~\ref{sec_results} are within a factor of two of the experimental measurements. The density obtained in simulations increases with increasing pressure, which is a trend that has also been observed in experiments. The important conclusions of the paper and suggestions for future work are presented in Sec.~\ref{sec_conclusions}.

\section{The Rigid-Beam Model}\label{rb_model}

The RB model is a simplification of both the e-beam dynamics and Maxwell’s equations that results in a set of equations that are easier and faster to solve but retain the gas breakdown physics driven by a rapidly rising, high-current-density electron-beam. The equations for the electric and magnetic fields, when coupled to dynamic equations for the plasma response and to a plasma chemistry model, provide a self-consistent set of equations for the evolution of the plasma and the fields. This model retains the essence of the gas breakdown problem and is an ideal platform to develop, compare, and contrast different plasma chemistry and response models.  In the remainder of this section, the reduction of Maxwell’s equations to a simpler set of equations is developed. 

A major approximation used to derive the RB model is that the net particle currents are assumed to be large compared to the displacement current so that $\epsilon_0 \partial \E/\partial t \ll \J_{\rm net}$, where $\J_{\rm net}$ is the net current density obtained by summing the primary beam current density $\J_{b}$, and the plasma current density $\J_{p}$. This approximation is good when the plasma density is large compared to the beam density so that the beam-plasma system is nearly charge neutral. Under these conditions, the electrostatic field is negligible and the inductive electric field associated with the rapidly rising beam current is the dominant field. The RB model therefore does not apply to the early time beam-plasma behavior before space-charge neutralization is established, or to subnanosecond pulses. It is assumed that this non-neutral phase is very short compared to the rise time of the beam so that the main electric field is the inductive electric field over the majority of the beam pulse. An order-of-magnitude estimate based on beam parameters produced in experiments\cite{ddh_beams} indicates that, at 1~Torr pressure, beam-impact ionization will charge-neutralize the beam in about 1~ns.

Ignoring displacement currents in Ampère’s law, the equations for the electric and magnetic fields can be written as
\begin{eqnarray}\label{faraday}
\frac{\partial \B}{\partial t} &=& -\curl \E, \qquad\qquad \text{(Faraday's Law)} \\
\curl \B &=& \mu_0 \left(\J_b +\J_p\right), \qquad \text{(Ampère's Law)} \label{ampere}
\end{eqnarray}
where SI units are used. Taking the curl of Faraday’s law and substituting Ampère’s law to eliminate the magnetic field allows these two equations to be combined into a single Poisson-type equation for the electric field which is given by
\begin{eqnarray}\label{eq:rb_field}
\nabla^2\E = \mu_0 \frac{\partial \J_b}{\partial t} + \mu_0 \frac{\partial \J_p}{\partial t},
\end{eqnarray}
where it has been assumed that $\nabla\cdot\E \simeq 0$. Equation \eqref{eq:rb_field} shows that the initial source of the inductive electric field is the rapidly rising, e-beam current density $\J_b$. Another important term that determines the overall magnitude and duration of the electric field is the plasma current density $\J_p$. During the rising part of the beam pulse, the plasma current density opposes the beam current, which tends to reduce both the net current and the magnitude of the electric field. During the falling part of the beam pulse, the plasma current is in the same direction as the beam current and can enhance the magnitude of the electric field.

It is often assumed that nearly complete current neutralization occurs when modeling the transport of high current-density electron beams in low pressure gas.\cite{beams_book} In this work, however, no assumptions are made about the magnitude of the induced plasma current $\J_p$. Rather, both the plasma current and the induced electric field are modeled using a set of coupled equations.

To simplify Eq.~\eqref{eq:rb_field} even further, it is assumed that the beam is cylindrically symmetric and flows in the positive z-direction. It is also assumed that spatial gradients in the direction of the beam propagation are small compared to radial gradients. 
This also implies that the pulse is long enough so that rapid temporal variations do not induce steep axial gradients.
With these assumptions, both the beam and plasma current densities can be written as \begin{eqnarray}
\J_b &=& J_b(r, t) {\bf e}_z, \label{rbmodel_jb}\\
\J_p &=& J_p(r, t) {\bf e}_z.\label{rbmodel_jp}
\end{eqnarray}
With these assumptions, Eq. \eqref{eq:rb_field} simplifies to
\begin{eqnarray}\label{eq:rb_field_1d}
\frac{1}{r}\frac{\partial}{\partial r} r \frac{\partial E_z}{\partial r} = \mu_0 \frac{\partial J_b}{\partial t} + \mu_0 \frac{\partial J_p}{\partial t}.
\end{eqnarray}
Once the beam and plasma current densities are known ($J_b$ specified and $J_p$ computed) and boundary conditions are specified, this equation can be solved for the electric field. The boundary condition at $r = 0$ is that $\partial E_z/{\partial r}|_{r=0}=0$ and, assuming a perfectly conducting wall at $R_w$, the other boundary condition is $E_z (R_w )=0$. The plasma model described in Sec.~\ref{plasma_response} is used to compute the plasma current density required to solve Eq.~\eqref{eq:rb_field_1d}. After solving Eq.~\eqref{eq:rb_field_1d} for $E_z$, Eq.~\eqref{faraday} can be used to calculate $B_\theta$.

\section{Plasma response models}\label{plasma_response}

The simplifications described in the previous section deal with how the e-beam and the electromagnetic fields are treated in the RB model. In addition to these approximations, a plasma response model is needed so that the plasma current can be computed for the source term in the RB model.

As the plasma density grows, the plasma current can become a significant fraction of the beam current. Since the plasma electrons are much more mobile than plasma ions, and the beam time scale is much shorter than the plasma hydrodynamic response time, ions will be treated as an immobile species that provides overall charge neutrality. Under this assumption, the plasma current density will be determined entirely from the motion of the plasma electrons. The plasma current density can then be written as
\begin{eqnarray}
J_p = - en_e V_e,
\end{eqnarray}
where $e$ is the magnitude of the electron charge, $n_e$ is the plasma electron density, and $V_e$ is the electron drift velocity that occurs as a result of acceleration in the electric field and scattering collisions. The electron density and the drift velocity are defined by moments of the electron energy distribution function, which itself depends on the electric field and the details of the plasma chemistry. In the remainder of this section, the details of the plasma chemistry and the fluid model used in the results section are presented.

\subsection{Plasma chemistry}
The plasma chemistry used in this paper includes any physical changes to the individual molecules such as excitation or ionization that result from energetic collisions between electrons and gas particles. This includes elastic scattering as well as inelastic collisions where a fraction of the electron energy is transferred to the molecule to produce rotational, vibrational, and electronic excitations, and ionization. Since the purpose of the present paper is to introduce the rigid-beam model as a way to study electron-beam-produced plasmas, for simplicity it will be assumed that the e-beam is injected into a gas consisting entirely of ground state N$_2$ molecules. The role of oxygen and other trace constituents of air (such as water vapor) in these plasmas will be the subject of future work. 

A complete plasma chemistry model describes processes that can take place between the plasma electrons and all heavy species, neutral and ionized, that may be present. In addition to collisions between thermal plasma electrons and the gas, collisions that result from the direct impact of the beam electrons and the gas can also be significant. In the remainder of this section, the electron-impact driven interactions that will be used in this paper are described. 

\subsection{The weakly ionized plasma model}

The weakly ionized plasma model is the simplest plasma chemistry model for an electrical discharge. In the weakly ionized model, only reactions between electrons and N$_2$ gas need to be followed. The possible end products of these reactions are either rotationally excited, vibrationally excited, or electronically excited molecules, or molecular ions. It is assumed that the densities of these excited or ionized species are small compared to the ground state neutral density so that the reactions between electrons and excited species can be neglected.  For weakly ionized N$_2$, the complete set of collisions can all be written in the form
\begin{eqnarray}\label{eqn_electron_collision}
{\rm e}+{\rm N}_2 \rightarrow  a {\rm e} + {\rm N}_2^* +\varepsilon^*,
\end{eqnarray}
where N$_2^*$ represents an excited or ionized state of nitrogen and $a$ is an integer that indicates how many electrons are produced in these reactions. For molecular nitrogen, $a=1$ for excitation processes (rotational, vibrational, and electronic excitation), and $a=2$ for ionization. For elastic collisions, where there is no change in internal energy of the molecule, $a=1$ and ${\rm N}_2^*={\rm N}_2$. The kinetic energy transfer between electrons and N$_2$ molecules during an elastic collision is small because of the small electron-to-molecule mass ratio. While elastic collisions are much more frequent than inelastic collisions, the loss of electron energy due to these collisions is less than 7\% of the inelastic energy loss for the rate tables used in this paper. When the mean electron energy is greater than 0.5~eV, the loss is even smaller, less than 0.1\%. The energy loss due to elastic collisions is therefore neglected in this paper.  Inelastic collisions result in quantized energy transfers between the plasma electrons and molecules, increasing the internal energy of the molecules by the amount $\varepsilon^*$. In addition to the reactions given in Eq. \eqref{eqn_electron_collision}, electrons with energies above the 9.75~eV dissociation threshold can drive reactions which result in the dissociation of the nitrogen molecule into two nitrogen atoms or ions. The atomic products of these dissociations are not tracked in this model.

In the weakly ionized plasma model, it assumed that, in general, the densities of all the excited  species that result from collisions with electrons is small so that reactions with these excited products can be neglected. However, for diagnostic purposes and to prepare for future versions of the model where the weakly ionized approximation breaks down, the density of the ground-state neutrals as well as the densities of all excited species are tracked using rate equations. The time-dependent rate equations for the type of reactions described by Eq. \eqref{eqn_electron_collision} are given by
\begin{eqnarray}
\label{eqrate1}\frac{dn_{{\rm N}_2}}{dt} &=& -n_{\rm e} \sum_j \nu_{{\rm N}^*_{2,j}}, \\
\label{eqrate2}\frac{dn_{{\rm N}^*_{2,j}}}{dt} &=& n_{\rm e}  \nu_{{\rm N}^*_{2,j}},
\end{eqnarray}
where $\nu_{{\rm N}^*_{2,j}} = n_{\rm N_2} k_{{\rm N}^*_{2,j}}$ is the collision frequency for the $j^{\rm th}$ excited state, and $k_{{\rm N}^*_{2,j}}$ is the rate coefficient obtained by averaging the energy-dependent cross sections over the electron energy distribution function. 
The density of N$_2$ at $t=0$ is set based on the pressure of the background gas, which is one of the main input parameters for this model. The densities of all other excited states at $t=0$ are set to zero. Beam impact ionization as well as dissociative recombination will be treated in Sec.~\ref{sec_bii} and will result in additional terms being added to Eqs.~\eqref{eqrate1} and \eqref{eqrate2}.

The rate coefficients are defined by
\begin{eqnarray}\label{eq_rate_coeff}
k_{{\rm N}^*_{2,j}} = \int_0^\infty d\varepsilon \, \sigma_{{\rm N}^*_{2,j}} (\varepsilon) v_\varepsilon f_0 (\varepsilon), 
\end{eqnarray}
where $\sigma_{{\rm N}^*_{2,j}} (\varepsilon)$ is the cross section for the reaction, $v_\varepsilon = \sqrt{2\varepsilon/m}$ is the speed of an electron at energy $\varepsilon$, and $f_0(\varepsilon)$ is the electron energy distribution function (eedf). The eedf is normalized so that
\begin{eqnarray}
\int_0^\infty d\varepsilon \, f_0(\varepsilon) = 1.
\end{eqnarray}
The time dynamics of the eedf is described by the Boltzmann equation. However, in this paper, we will use a simplification where the eedf takes the form of a Maxwellian distribution with a time-dependent temperature. Further details will be described in Sec.~\ref{sec_thermal_electrons}.

Each type of collision is represented by a cross section and a threshold energy. The N$_2$ collisional cross section data used in this paper is the Phelps data set downloaded from the LXCat website.\cite{phelps_lxcat} The individual cross sections in this database come from a number of sources which can be found in Ref.~\onlinecite{PhysRevA.31.2932}. In this data set, there are a total of 24 different collision types with each identified by the excited product that results from the collision. The complete set of reactions and thresholds is summarized in Table~\ref{tab_plasma_chemistry}. Molecular term notation is used to describe both the ground state (X$\,^1\Sigma$) of the N$_2$ molecule and the electronic excited states. There is one momentum transfer collisional cross section and one for the aggregate rotational mode of X$\,^1\Sigma$ labeled as {\em rot}. There are cross sections for eight vibrational modes of X$\,^1\Sigma$ with vibrational quantum numbers $v=1\ldots 8$, for thirteen molecular electronic excited states, and there is one total ionization cross section. Even though the ionization reaction includes dissociative ionization events as well as ionization into excited states of N$_2^+$ such as A$\,^2\Pi$ and B$\,^2\Sigma$, the ionization product state is labeled in this paper by N$_2^+$. The electronic triplet state A$\,^3\Sigma$ is divided into three lumped vibrational bands as indicated in the table. No vibrational resolution is included for any of the other electronic states. 

In Refs.~\onlinecite{PhysRevA.31.2932} and \onlinecite{phelps_lxcat} the authors include a cross section for excitation to a sum of singlet states (labeled {\em sum} in Table~\ref{tab_plasma_chemistry}).
The singlet states have threshold energies between 12.5 and 14.8~eV which are well above the 9.75~eV dissociation energy for the nitrogen molecule. As a result, the excited singlet states are highly unstable and often result in the dissociation into two nitrogen atoms. In Ref.~\onlinecite{ZIPF1978449}, the fraction of the excited singlet sum states that dissociates is estimated to be 0.73. This dissociation pathway as well as the pathway through high lying triplet states is ignored. Future work will consider these pathways for regimes where this is important.

Collisions between electrons and excited molecular states of nitrogen (momentum transfer, step-wise excitation and step-wise ionization) as well as collisions between electrons and atomic nitrogen are beyond the scope of the weakly ionized plasma assumption. Some of these processes were considered in Ref.~\onlinecite{doi:10.1063/1.4950840}, and they will also be the subject of future papers. There are many additional processes which affect the excited state populations but are not included in the plasma chemistry model used in this paper. These neglected processes include collisional de-excitation, spontaneous radiative de-excitation, and heavy-heavy reactions such as charge exchange. Future work should include the physics of these processes in order to more accurately track the densities of the excited states.

\renewcommand{\arraystretch}{1.1}
\begin{table*}
\caption{\label{tab_plasma_chemistry}A summary of the weakly ionized plasma model for N$_2$.}
\begin{ruledtabular}
\begin{tabular}{lcccc}
Collision Type & Products & Threshold [eV] & \shortstack{Beam-impact\\Production Efficiency $\xi_{{\rm N}_{2,j}^*}$}  \\
\hline
Elastic		&  X$\,^1\Sigma$ & 0 & --- \\[2pt]
Rotational\footnotemark[1]	& {\em rot} & 0.02 & 20.25 \\[4pt]
Vibrational\footnotemark[1]	&  $v=1$, 2, 3, 4 
		&  0.29, 0.59, 0.88, 1.17 
		&  5.237, 0.391, 0.147, 0.104  \\
		&  $v=5$, 6, 7, 8 
		&  1.47, 1.76, 2.06, 2.35 
		&  0.100, 0.075, 0.070, 0.049\\[4pt]
Electronic	&
A$\,^3\Sigma$($v={0-4}$), A$\,^3\Sigma$($v={5-9}$) 
& 6.17, 7.0 
& 0.167/3\footnotemark[2], 0.167/3\footnotemark[2]\\
&
B$\,^3\Pi$, W$\,^3\Delta$, A$\,^3\Sigma$($v\ge{10}$), B$^\prime\,^3\Sigma$ 
& 7.35, 7.36, 7.8, 8.16 
& 0.148, 0.112, 0.167/3\footnotemark[2], 0.030\\
&a$^\prime\,^1\Sigma$, a$\,^1\Pi$, w$\,^1\Delta$, C$\,^3\Pi$ 
& 8.4, 8.55, 8.89, 11.03 
& 0.028, 0.081, 0.031, 0.057\\
& E$\,^3\Sigma$, a$^{\prime\prime}\,^1\Sigma$, {\em sum} 
& 11.87, 12.25, 13.0 
& 0.001, 0.010, 0.664
\\[4pt]
Ionization	&  N$_2^+$ & 15.6 & 1 \\[4pt]
Dissociative Recombination\footnotemark[3] & Atomic N & 0 & --- 
\end{tabular}
\footnotetext[1]{These are rotational and vibrational excitations of the ground state, N$_2$(X$\,^1\Sigma$).}
\footnotetext[2]{The beam-impact excitation of the A$\,^3\Sigma$ state is not vibrationally resolved in Ref.~\onlinecite{doi:10.1063/1.345772}. In this model 1/3 of the excitations are attributed to each of the three vibrational bands.}
\footnotetext[3]{Unlike the other reactions in this table, the target species for this reaction is N$_2^+$.}
\end{ruledtabular}
\end{table*}

There is one additional reaction beyond those described above that is tracked in this paper, and that is dissociative recombination of molecular nitrogen ions:
\begin{eqnarray}
{\rm e} + {\rm N}_2^+ \rightarrow 2 {\rm N}.
\end{eqnarray}
For this reaction, we use the rate constant from Ref.~\onlinecite{doi:10.1063/1.475577} (which was also used in Ref.~\onlinecite{doi:10.1063/1.4950840}),
\begin{eqnarray}\label{dr_rate}
k_{dr} = 5.85 \times10^{-8}/ T_e^{0.30}  \, {\rm cm}^3/{\rm s}, 
\end{eqnarray}
where the electron temperature $T_e$ is in eV. Including this reaction provides an estimate for the production of atomic nitrogen via this pathway, and it also captures the reduction in plasma density due to recombination. This is the only reaction in this plasma chemistry model which allows for the decay of electron density, a phenomenon which is experimentally observed.
With this reaction we also have a rate equation for atomic nitrogen:
\begin{eqnarray}
\frac{dn_\text{N}}{dt} = 2n_e n_{\text{N}_2^+} k_{dr}.
\end{eqnarray}
The factor of two on the right hand side comes from the fact that the dissociation of one molecular ion produces two atoms of nitrogen.
Loss terms due to dissociative recombination are also added to the electron and ion rate equations.
Note that since the ion density $n_{\text{N}_2^+}$ is equal to the electron density in this quasi-neutral model, the source term for atomic nitrogen, and the associated loss terms in the electron and ion rate equations, are second order in $n_e$. All other processes considered in this paper are first order in $n_e$.  

\subsection{Plasma dynamics}
The way that the plasma chemistry is used depends largely on the model used for the plasma dynamics. For example, in a kinetic model, the plasma chemistry is represented by a set of energy-dependent collisional cross sections. In a fluid model, a set of collisional rates is determined by Eq.~\eqref{eq_rate_coeff} with assumptions about the electron energy distribution function. The fluid model described in this section will be used to describe the plasma dynamics in the remainder of this paper. 

\subsubsection{The fluid equations for the thermal plasma}\label{fluid_model}

The plasma electrons can be treated as a fluid provided the collision rate is sufficiently large such that the mean-free time between collisions is small compared to the rise time of the beam, and the mean-free path is short compared to the length scales in the problem. If $\V^\prime$ and $U^\prime$ are the mean velocity and energy of the secondary electrons created from ionization events by thermal electrons, and $\V_s$ and $U_s$ are the mean velocity and energy of secondary electrons from beam-impact ionization, then the thermal electron fluid equations for the beam-plasma system in conservative form are given by
\begin{eqnarray}
\label{fluideq1}\frac{\partial n_e}{\partial t} &&+ \nabla \cdot (n_e \V_e) = \nonumber\\
&&\left.\frac{dn_e}{dt}\right|_{\rm beam} + n_e \left( \nu_{\text{N}_2^+} -  n_{\text{N}_2^+} k_{dr}\right), \\
\frac{\partial n_em\V_e}{\partial t} &&+ \nabla \cdot (n_e m \V_e \V_e) = -e n_e (\E + \V_e\times \B) -\nabla p_e \nonumber\\
&& - \nu_m n_e m \V_e  + \left.\frac{dn_e}{dt}\right|_{\rm beam} m \V_s,  \nonumber\\
&& + n_em\left(\nu_{\text{N}_2^+}  \V^\prime -n_{\text{N}_2^+}k_{dr}\V_e \right) \label{fluideq2}
\\
\frac{\partial n_e U_e}{\partial t} &&+ \nabla \cdot (n_e U_e \V_e) = -en_e\V_e\cdot\E - \nabla\cdot p_e \V_e \nonumber\\
&& - \nu_\varepsilon n_eU_e  +\left.\frac{dn_e}{dt}\right|_{\rm beam} U_s \nonumber\\
&& + n_e \left(\nu_{\text{N}_2^+} U^\prime -n_{\text{N}_2^+}k_{dr}U_e\right), \label{fluideq3}
\end{eqnarray}
where $p_e$ is the plasma electron pressure, $\nu_{\text{N}_2^+}$ is the total thermal electron ionization rate, $\nu_m$ is the momentum transfer frequency, $\nu_\varepsilon$ is inelastic energy exchange frequency, and $\left.\frac{dn_e}{dt}\right|_{\rm beam}$ is the beam-impact ionization rate. The last two terms on the right-hand sides of each equation represent the volumetric rates for adding or removing electron density, momentum, and energy during ionization and recombination events.
To close the equations, the pressure must be given in terms of the fluid variables, or it can be neglected if appropriate for the parameters of interest (Sec.~\ref{sec_thermal_electrons}). 

\subsubsection{Beam-impact ionization}\label{sec_bii}
Ionization of neutral gas by energetic beam particles plays a significant role in the overall plasma evolution, especially at higher gas pressure. 
Early in time, the energetic beam electrons collisionally ionize the background gas and start to generate a  plasma. 
This plasma produced by the electron beam also provides seed electrons for thermal excitation and ionization processes.

The fluid model in this paper specifies the beam-impact ionization rate as well as the average momentum and energy of the secondary electrons from beam impact events. These quantities are needed for the source terms in the electron fluid equations. One common model is to represent the beam impact rate by
\begin{eqnarray}\label{bii}
\left.\frac{dn_e}{dt}\right|_{\rm beam} = \frac{|\J_b|}{eW}\rho_G S,
\end{eqnarray}
where $\J_b$ is the beam current density, $\rho_G$ is the gas density in g/cm$^3$, $S$ is the beam-electron-energy-dependent stopping power in eV-cm$^2$/g, and $W$ is the average energy expended by the beam per electron-ion pair.\cite{doi:10.1063/1.345772} The stopping power approach assumes that energetic secondary electrons produced by individual beam-impact events both excite the gas and produce additional electron-ion pairs as they slow down and join the thermal population. A detailed analysis for N$_2$ gas shows that the value of $W \simeq 36 \text{ eV}$ is nearly constant for beam energies above 1~keV.\cite{doi:10.1063/1.345772} The value of $W$ is quite a bit higher than the ionization energy since there are many competing inelastic excitation processes which act as energy sinks as the energetic secondary electrons produced by the beam slow down. Since Eq.~\eqref{bii} assumes that secondary electrons created by beam impact rapidly join the thermal population, it is reasonable to take $\V_s=\V_e$ and $U_s=U_e$ in Eqs.~\eqref{fluideq2} and \eqref{fluideq3}.

To solve Eq.~\eqref{bii}, the stopping power of electrons in nitrogen is needed. The total stopping power in N$_2$ (collisional plus radiative bremsstrahlung losses) can be calculated by the NIST electron stopping power program ESTAR.\cite{estar} The Bethe-Bloch formula\cite{SELTZER19821189} is an approximate expression for the stopping power which agrees well with the collisional part of the ESTAR stopping power below 100 MeV in nitrogen. Since this paper is concerned with e-beam energies below 100 keV, radiative stopping can safely be ignored in our analyses, and the Bethe-Bloch formula for stopping power from Ref.~\onlinecite{SELTZER19821189} is used.

The beam-impact ionization rate in Eq.~\eqref{bii} is valid for the case where the beam electrons are monoenergetic and have no spread in velocity (which could be due, for example, to scattering of the beam electrons in the foil when the enter the gas chamber). A spread in either energy or velocity could change both the stopping power and the effective path length of beam electrons. These effects are beyond the scope of this work.

In addition to ionization events, the beam can directly excite the gas. For N$_2$ gas, the number of excited species produced by the e-beam per electron ion pair is relatively constant for beam energies above 1~keV.\cite{doi:10.1063/1.345772} The number of excited species of type $j$ created per electron-ion pair is called the production efficiency, which is denoted $\xi_{{\rm N}_{2,j}^*}$.
A term is added to the rate equations for the neutral species to account for these beam-impact reactions:
\begin{eqnarray}\label{neutral_rate_1}
\frac{dn_{{\rm N}_2}}{dt} &=& -n_{\rm e} \sum_j \nu_{{\rm N}^*_{2,j}}
	-\left(1 + \sum_j \xi_{{\rm N}_{2,j}^*} \right) \frac{|\J_b|}{eW}\rho_G S, \\
\frac{dn_{{\rm N}^*_{2,j}}}{dt} &=& n_{\rm e}  \nu_{{\rm N}^*_{2,j}}
	+ \xi_{{\rm N}_{2,j}^*} \frac{|\J_b|}{eW}\rho_G S.  \label{neutral_rate_2}
\end{eqnarray}
The beam excitation production efficiencies are taken from Ref.~\onlinecite{doi:10.1063/1.345772} for a 1~keV beam, and modified slightly to be compatible with the Phelps cross section set that is used to compute thermal rates.  

The single Phelps ionization cross section is a lumped cross section which includes all final states of the N$_2^+$ ion as well as dissociative ionization to  N$^+$ whereas each of these ionization channels are tracked separately in Ref.~\onlinecite{doi:10.1063/1.345772}. Instead of tracking these final states separately, we sum all of the ionization channels from Ref.~\onlinecite{doi:10.1063/1.345772} to obtain a single beam-impact ionization production efficiency, with the final state being a generic ``ion''. In our model, this product species is added to the ${\rm N}_2^+$ species, which now represents a generic ``ion'' final state. 

Two other modifications are also made. First, the triplet state A$\,^3\Sigma$ in the Phelps dataset is vibrationally resolved in three bands, while the A$\,^3\Sigma$ production efficiency in Ref.~\onlinecite{doi:10.1063/1.345772} is not vibrationally resolved. In this paper, 1/3 of the beam-driven excitation of the A$\,^3\Sigma$ state is assigned to each of the three vibrational bands.
Second, the dissociation channels in Ref.~\onlinecite{doi:10.1063/1.345772} must be reconciled with the sum-of-singlets state that appears in the Phelps dataset. Both of these processes are consistent with the dissociation cross section in Ref.~\onlinecite{ZIPF1978449}, with the Phelps sum-of-singlets-state cross section being larger than that dissociation cross section by (1/0.73), where 0.73 is the dissociation fraction in Ref.~\onlinecite{ZIPF1978449}. Therefore, in our model, we take the total beam-impact dissociation efficiency from Ref.~\onlinecite{doi:10.1063/1.345772}, divide by the dissociation fraction of 0.73 from Ref.~\onlinecite{ZIPF1978449}, and attribute the result to the Phelps sum-of-singlets-state. The production efficiencies used in our model are summarized in Table~\ref{tab_plasma_chemistry}.

\subsubsection{Collisional frequencies}\label{sec_thermal_electrons}

As mentioned above, the fluid equations \eqref{fluideq1} -- \eqref{fluideq3} need frequencies that describe the interaction of the thermal electrons with the background gas. There are four frequencies for thermal interactions that need to be specified for the electron fluid: the ionization frequency, the momentum transfer frequency, the total inelastic energy transfer frequency, and the dissociation frequency. The dissociation frequency is $\nu_{dr} = n_{\text{N}_2^+}k_{dr}$, where the dissociative rate constant $k_{dr}$ is given in \eqref{dr_rate}.

The collisional ionization frequency by the thermal electrons is given by 
\begin{eqnarray}
\nu_{\text{N}_2^+} = n_{\text{N}_2} k_{\text{N}_2^+} ,
\end{eqnarray}
where $k_{\text{N}_2^+}$ is the rate coefficient defined in Eq.~\eqref{eq_rate_coeff}. 

 The electron momentum equation requires the momentum transfer frequency for collisions between thermal electrons and neutral gas molecules. This frequency is defined by
\begin{eqnarray}
\nu_m = n_{\text{N}_2} \int_0^\infty d\varepsilon \, \sigma_{m} (\varepsilon) v_\varepsilon f_0(\varepsilon) ,
\end{eqnarray}
where $\sigma_m$ is the total momentum transfer cross section. This cross section is from all processes including elastic collisions, ionizing collisions, and all other inelastic collision processes. 

In the energy equation, an expression for the inelastic energy transfer between thermal electrons and gas molecules is needed. The inelastic energy-transfer frequency is defined by
\begin{eqnarray}
\nu_\varepsilon = \frac{1}{U_e} \sum_{ j } \varepsilon_{ {\rm N}_{2,j}^* } \nu_{ {\rm N}_{2,j}^* },
\end{eqnarray}
where $\varepsilon_{ {\rm N}_{2,j}^* }$ is the threshold energy and $\nu_{ {\rm N}_{2,j}^* }$ is the inelastic collision frequency for the process with product N$_{2,j}^*$. The sum is over all inelastic collision processes. The inelastic collision frequency is given by
\begin{eqnarray}
\nu_{ {\rm N}_{2,j}^* } = n_{\text{N}_2} k_{{\rm N}_{2,j}^*},
\end{eqnarray}
where $k_{{\rm N}_{2,j}^*}$ is the rate coefficient for producing the excited state N$_{2,j}^*$.

\begin{figure}
\includegraphics[width=\columnwidth]{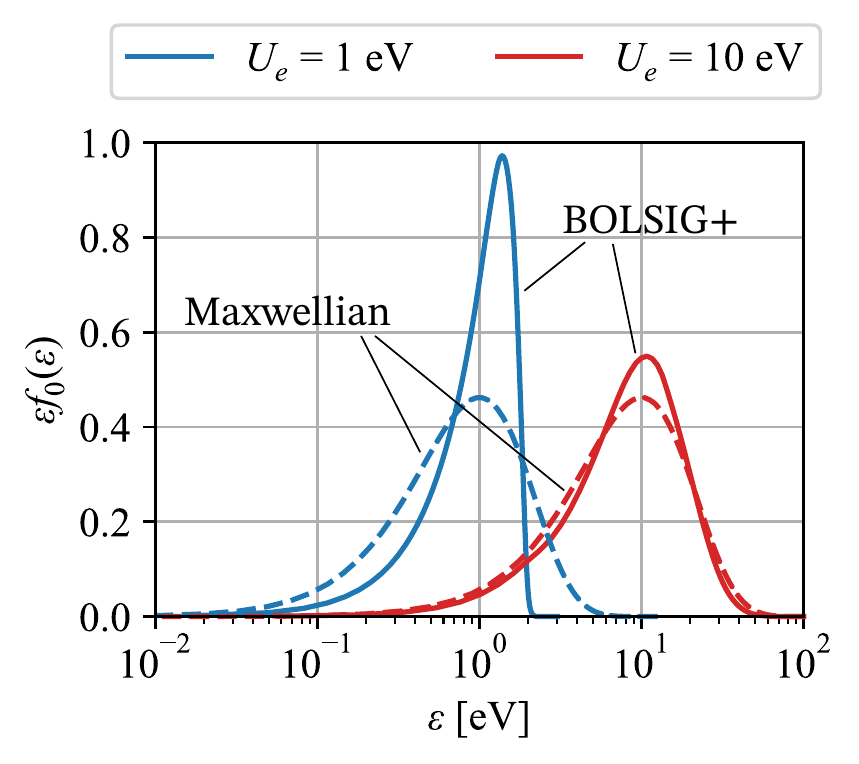}
\caption{\label{fig_eedf} A comparison of the steady-state electron energy distribution function computed using the BOLSIG+ code under the assumption of a weakly ionized plasma (solid) with a Maxwellian distribution (dashed).
}
\end{figure}

These frequencies cannot be evaluated without an estimate of the eedf, $f_0$. If the ionization fraction is low and Coulomb collisions are infrequent, the dynamics of the eedf is dominated by the energy transfer between the electrons and the neutral gas molecules via inelastic collision processes. In this case, a steady-state two-term approximation for the eedf $f_0(\varepsilon)$ can be used to calculate the collision frequencies. A code like BOLSIG+ can be used to calculate the the eedf in this approximation.\cite{Hagelaar_2005} If the ionization fraction is high and Coulomb collisions between electrons and other free electrons in the plasma dominate the total inelastic collision frequency, then the Coulomb collisions in the plasma will drive the eedf to a Maxwellian distribution. In this case, the eedf is
\begin{eqnarray}
f_0(\varepsilon) = 2 \varepsilon^{1/2} \frac{\exp(-\varepsilon/T_e)}{\sqrt{\pi} T_e^{3/2}},
\end{eqnarray}
where $T_e = \frac{2}{3} U_e$ is the electron temperature. Figure~\ref{fig_eedf} shows the differences between the steady state eedf computed by BOLSIG+ and a Maxwellian eedf, for two different values of the mean electron energy $U_e$. At high mean energy ($U_e$=10~eV, shown in red) there is not much difference between the two eedfs and it is valid to approximate the eedf by a Maxwellian. At low mean energy ($U_e$=1~eV, shown in blue), the steady-state eedf differs significantly from Maxwellian. Energy losses during inelastic collisions cause much of this difference. If there are no electron-electron Coulomb collisions to cause $f_0(\varepsilon)$ to spread in energy and relax to a Maxwellian, then the shape of $f_0(\varepsilon)$ is set by a balance between acceleration of electrons in the electric field and energy loss due to inelastic collisions.

\begin{figure}
\includegraphics[width=\columnwidth]{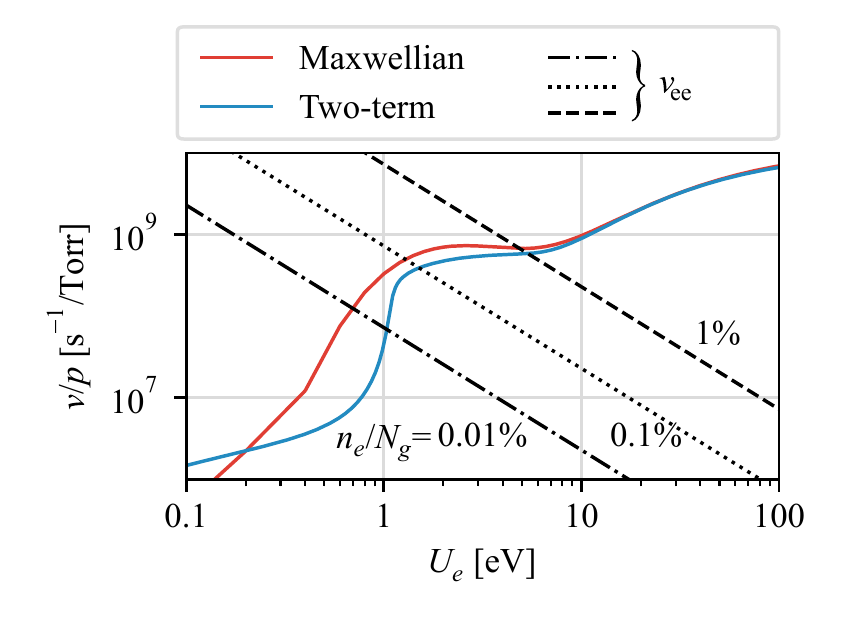}
\caption{\label{frequency_comparison}Comparison of inelastic collision frequency (red and blue) and estimates of the electron-electron collision frequency (black).
}
\end{figure}

In order to decide which approximate eedf to use in this work, a comparison can be made between the inelastic collision frequency and an estimate of the electron-electron collision frequency. In cgs units (but with temperature in eV), this collision frequency is:\cite{formulary}
\begin{eqnarray}\label{nu_ei}
\sqrt{2} \, \nu_\text{ee} \sim \nu_\text{ei} = 2.91\times10^{-6}  n_{e} \ln\Lambda T_e^{-3/2} .
\end{eqnarray}
Figure~\ref{frequency_comparison} shows a log-log plot of the inelastic collision frequency computed using a Maxwellian eedf (red), the inelastic collision frequency computed using a steady-state eedf (blue), and estimates of $\nu_\text{ee}$ made using ionization fractions of $n_e/N_g$ = 0.01\%, 0.1\%, and 1\%, and $\ln\Lambda = 10$ (dashed black lines), where $N_g$ is the number density of the background neutral gas. As this plot shows, electron-electron collisions occur much more frequently than inelastic collisions for $U_e \lesssim 1\text{ eV}$, even for plasmas with low ionization fraction. 
The electron-electron collisions drive the eedf towards a Maxwellian distribution, while the inelastic collisions drive it away from Maxwellian. In the low energy region where the Maxwellian and two-term collision frequencies differ more, the electron-electron collisions occur more often than the inelastic collisions, and we would expect them to drive the eedf towards a Maxwellian.
Because of this, we will use collision rates based on a Maxwellian eedf assumption. More accurate collision frequencies could be computed by coupling this model to a time-dependent solver for the eedf, however this is beyond the scope of this paper.

\begin{figure}
\includegraphics[width=\columnwidth]{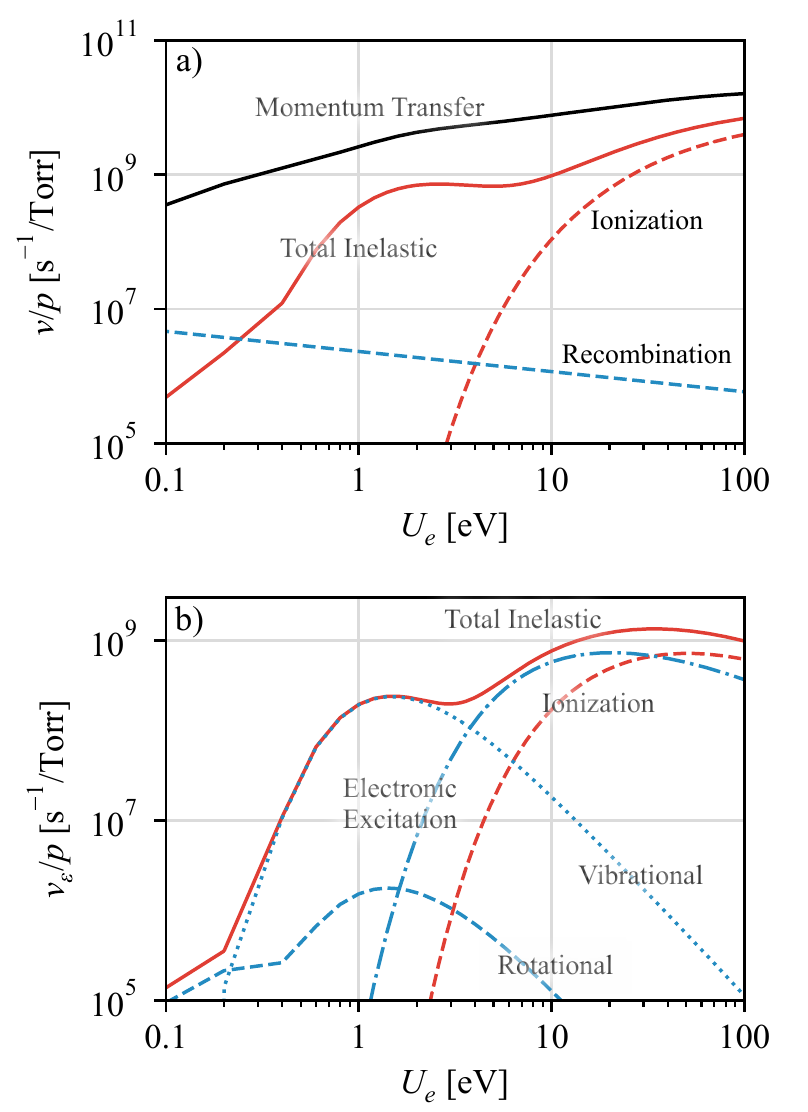}
\caption{\label{fig_rates} The collision frequencies for the plasma chemistry used in this paper, computed using a Maxwellian eedf.  The total collision frequencies are shown in a) along with the recombination frequency for $n_e/N_g=0.1$\%. The energy-transfer frequencies are shown in b). The dashed curves in b) show the contribution from various inelastic processes to the total inelastic energy transfer frequency.
}
\end{figure}

Figure~\ref{fig_rates} shows a summary of the collision frequencies used in this work computed using a Maxwellian eedf. The recombination frequency in Fig.~\ref{fig_rates}(a) is shown for an ionization fraction of $n_e/N_g=0.1$\%. The total inelastic energy transfer frequency shown in Fig.~\ref{fig_rates}(b) as a solid red line is the sum of contributions from all of the inelastic processes. The dashed lines show how the various types of process contribute to the inelastic energy transfer frequency. Since energy transfer due to elastic collisions between electrons and the neutral molecules is reduced by their mass ratio, it is very small compared to inelastic energy transfer and is neglected in this work. Note that the ionization frequency is nonzero even for mean energies below the ionization threshold. This is due to the tail of the eedf extending above the threshold, even when the mean energy is below threshold.

\subsubsection{Simplified fluid equations}
To proceed with the fluid equations, it is also necessary to make some assumptions about the average velocity and energy for secondary electrons produced during ionization events by thermal electrons. It is common to assume that secondary electrons are produced with an isotropic distribution. With this assumption, the average secondary velocity is given by $\V^\prime=0$. Since momentum is conserved during the ionizing collision, any momentum lost by the primary electron must show up in the ion. Since $\nu_m$ represents the momentum transfer for both inelastic and elastic collisions, the momentum transfer to ions is already included. Ignoring any kinetic energy in the ion, the average secondary kinetic energy is approximated by $U^\prime=U_e$ in the electron energy equation. Energy lost to ionization is already one of the terms in the inelastic energy loss. It is also assumed that $\V_s=\V_e$ and $U_s=U_e$.

Because of the symmetry assumptions in the rigid-beam model, Eqs.~\eqref{rbmodel_jb} and \eqref{rbmodel_jp}, the fluid equations simplify because the divergence terms on the left-hand sides of Eqs.~\eqref{fluideq1} -- \eqref{fluideq3} are zero. Another simplification comes from comparing the $\V_e\times\B$ term with the term due to momentum transfer with the neutrals. The ratio of these terms is proportional to the Hall coefficient, C$_\text{H} = \omega_\text{ce}/\nu_m$, where the electron cyclotron frequency is $\omega_\text{ce} = eB/m_e$. For the parameters being considered, the Hall coefficient is fairly small, and so the $\V_e\times\B$ term will be neglected. However, at lower pressures, the Hall coefficient can reach a maximum value greater than one, as will be discussed in Sec.~\ref{sims_vs_pressure}. This is an indication that, at lower pressures, the magnetic field and the radial component of the electric field may need to be included.

The pressure gradient term in Eq.~\eqref{fluideq2} will also be neglected in this model. This approximation is justified because the pressure gradient gives rise to a radial electric field, which holds the electrons back and accelerates the ions outward radially. The resulting expansion of the plasma happens at the ion sound speed, which is small for the parameters of this plasma.

With these simplifications, the fluid equations \eqref{fluideq1} -- \eqref{fluideq3} for the plasma electrons become:
\begin{eqnarray}
\frac{\partial n_e}{\partial t} &=& n_e\left(\nu_{\text{N}_2^+}  - n_{\text{N}_2^+} k_{dr}\right) + \left.\frac{dn_e}{dt}\right|_{\text{beam}}, \label{simp_fluid_1}\\
\frac{\partial n_emV_{z,e}}{\partial t} &=& -en_eE_z - \nu_\text{eff} n_emV_{z,e}\label{simp_fluid_2}\\
\frac{\partial n_eU_e}{\partial t} &=& -en_eV_{z,e}E_z - (\nu_\varepsilon - \nu_{\text{N}_2^+})n_eU_e \nonumber\\
&&+ U_e\left.\frac{dn_e}{dt}\right|_{\text{beam}} 
-n_en_{\text{N}_2^+} k_{dr} U_e
. \label{electron_energy_eqn}
\end{eqnarray}
where we have defined the effective momentum transfer frequency as
\begin{eqnarray}\label{eff_mom}
\nu_\text{eff} \equiv \nu_m + n_{\text{N}_2^+} k_{dr} - \frac{1}{n_e}\left.\frac{dn}{dt}\right|_{\text{beam}}.
\end{eqnarray}

Note that only the $z$ component of the momentum equation \eqref{simp_fluid_2} remains,
and that it can be used to obtain an equation for the plasma current:
\begin{eqnarray}
\frac{\partial J_{p}}{\partial t} = \nu_\text{eff} (\sigma E_z-J_{p}) .
\end{eqnarray}
In this equation, we have used the definition of the conductivity
\begin{eqnarray}
\sigma \equiv \frac{e^2n}{m\nu_\text{eff}}.
\end{eqnarray}
The second two terms on the right-hand side of Eq.~\eqref{eff_mom} are orders of magnitude smaller than the momentum transfer frequency $\nu_m$ for the parameters of our beam. 
Early in the beam pulse, the beam impact ionization term can be larger than the other terms but it quickly becomes small as the electron density rises.
Because these terms are small, they will be neglected in the remainder of this paper, and we will assume that $\nu_\text{eff} \simeq \nu_m$.

These fluid equations \eqref{simp_fluid_1}--\eqref{electron_energy_eqn}, together with the rate equations \eqref{neutral_rate_1} and \eqref{neutral_rate_2} for the heavy species and equation \eqref{eq:rb_field_1d} for the electric field, form the set of equations that are solved numerically for this problem.

\section{Numerical simulations}\label{sec_results}
The equations for the rigid-beam model, together with the fluid response model for the plasma, have been solved numerically for parameters selected to match experiments that were performed at NRL. In these experiments, a pulsed electron beam is injected into a cylindrical chamber filled with low-pressure gas. Experimental diagnostics measure various aspects of the plasma response, including the net current, the line-integrated plasma density at several locations, and framing camera images and spectra of the light emitted by the plasma. The pulse length of the electron beam is about 100~ns, the peak current in the pulse is about 4~kA, and the beam diameter is about 4~cm full width at half maximum. The pressure in the gas chamber ranges from about 100~mTorr to 10~Torr. Details of the experiments and analysis of the experimental results were reported elsewhere.\cite{ddh_beams}

\begin{figure}
\includegraphics[width=\columnwidth]{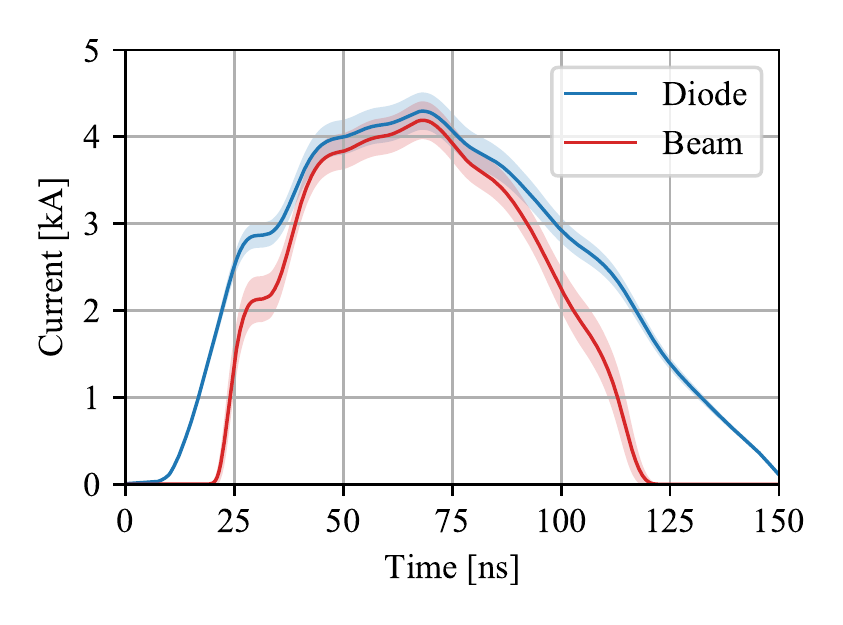}
\caption{\label{beam}Diode current (blue) and beam current (red) histories. Shaded regions represent experimental uncertainty in the diode current. The beam current is lower than the diode current because of current lost in the foils between the diode and the gas chamber.
}
\end{figure}

\begin{figure}
\includegraphics[width=\columnwidth]{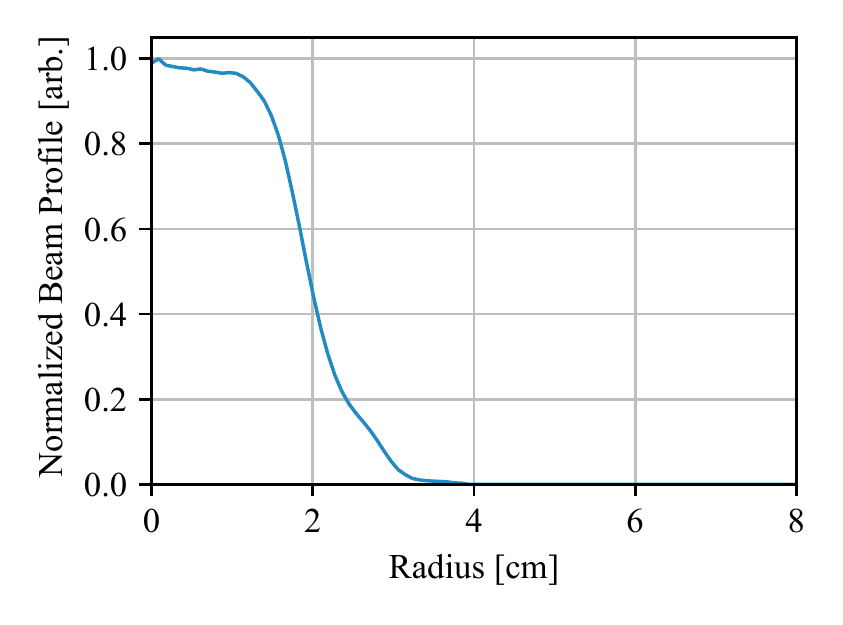}
\caption{\label{beam_profile}Normalized radial profile of the electron beam.}
\end{figure}

The simulations reported in this section are driven with beam current and voltage histories derived from experimental measurements. The time variations of the diode current are shown in Fig.~\ref{beam} (blue curve), with a blue shaded region that represents experimental measurement uncertainty. Shot-to-shot variations in the diode current give a smaller uncertainty, and are thus neglected in this work. 
Electrons are extracted from the diode and injected into a gas-fill transport chamber through a pair of 8.52-$\mu$m-thick aluminized Kapton foils. The beam is attenuated and scattered by the foils which results in a reduction of both the energy and current that the beam has as it enters the transport section. 
 Monte Carlo simulations of the current loss provided an estimate of the beam current entering the gas cell. This current is shown in red in Fig.~\ref{beam}, along with a shaded region that represents the uncertainty in the diode current measurement. The normalized radial profile of the beam is inferred from experimental measurements, and shown in Fig.~\ref{beam_profile}.

\begin{figure}
\includegraphics[width=\columnwidth]{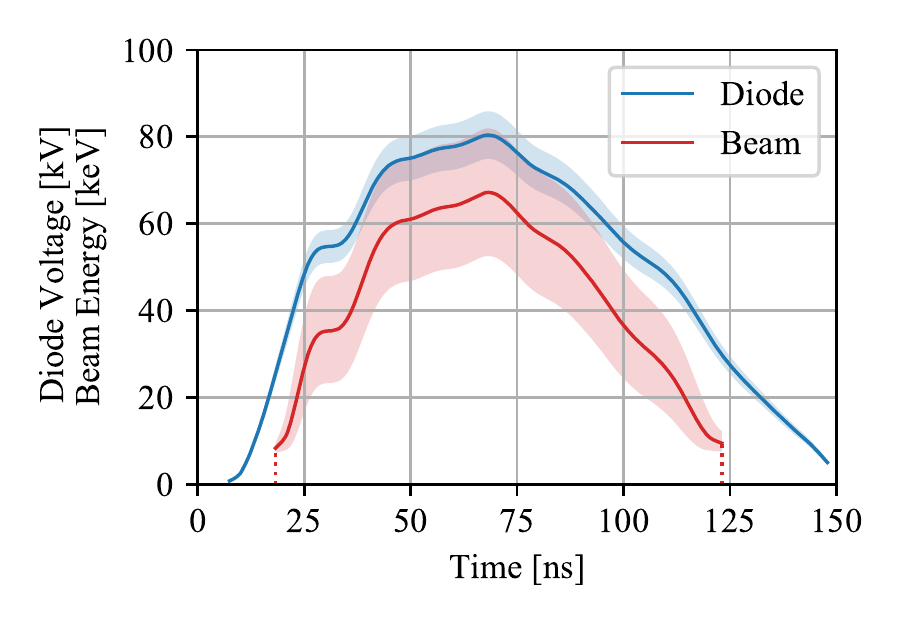}
\caption{\label{beam_voltage}Diode voltage (blue) and beam electron energy (red) histories. The gray shaded region represents experimental uncertainty in the diode voltage. The blue shaded region represents both experimental voltage uncertainty and spread in electron energy due to scatter in the foils. At low voltage, the beam is completely stopped by the foil.
}
\end{figure}

The energy of the beam electrons is estimated from the measured diode voltage, which is shown in Fig.~\ref{beam_voltage} (blue curve) along with an estimate of the uncertainty in the voltage measurement (blue shaded region). The Monte Carlo calculations described above also give an estimate of the scattering and slowing of the beam electrons in the foils. The statistical nature of the energy loss in the foils introduces a spread in the energy of the electrons that enter the gas chamber. The red curve in Fig.~\ref{beam_voltage} shows the mean energy of these beam electrons. 
The red shaded region comes from propagating the uncertainty in the measured diode voltage through the Monte-Carlo simulations. Since it is calculated using nonlinear uncertainty propagation in this way, it accounts for both the energy spread induced by electron scattering as well as the experimental uncertainty in the voltage measurements. When the diode voltage is low, the electrons are completely stopped in the foil. This leads to a narrower beam pulse as compared to the diode current, which can be seen in Fig.~\ref{beam}.

In the simulation results that are reported in this section, estimates of uncertainty in the simulation outputs are obtained by running a series of simulations with different combinations of beam histories. For beam current (and beam energy) there are three options: the mean value, the upper uncertainty bound, and the lower uncertainty bound. Combining these options gives a total of nine simulations that are run in order to get a simulation result with an estimated uncertainty. In the plots of the simulation results shown in this section, the mean of these nine results will be shown with a solid line, and the standard deviation will be shown using shaded uncertainty bounds.

Note that there are additional parameters of the simulations (such as tabulated collision rates, gas pressure, initial electron density, or beam radial profile) that have experimental uncertainties associated with them. These uncertainties are not accounted for in the simulations presented here. Only the uncertainties in the beam current and energy have been propagated through the simulation. A more complete and systematic investigation of these other uncertainties is beyond the scope of this work.

\subsection{Custom turboPy modules for the rigid-beam model}

To quickly implement and test the various parts of the rigid-beam model, we use the Python physics simulation framework turboPy.\cite{turbopy_code,turbopy_code_v2020.10.14} This framework manages boilerplate aspects of the code such as file I/O and problem configuration, while also providing basic computational physics functionality such as defining the main simulation loop, a clock, and a grid.\cite{RICHARDSON2020107607} For this rigid-beam problem, we have created custom turboPy PhysicsModules which solve the electric field equation, the rate equations, and the electron fluid equations. Additional PhysicsModules read in tabulated rate coefficients and perform the rate table lookups, while others calculate the stopping power and beam-impact ionization and excitation rates. In addition to these Physics\-Modules, several custom Diagnostics were written to compute quantities such as the line-integrated electron density (for comparison with experiment) and the electron-electron collision frequency (for checking the validity of the weakly ionized model). While most of these PhysicsModules use standard numerical techniques, there is one whose details are worth describing. 

The equations for the electric field and the plasma current are closely coupled in this model. In particular, the electric field equation \eqref{eq:rb_field_1d} does not have the form of a temporal evolution equation, but rather is a Poisson-like second-order differential equation that must be inverted to find $E_z$. In the code this is done by combining it with the equation for the plasma current to obtain
\begin{eqnarray}
\left(\frac{1}{r}\frac{\partial}{\partial r} r \frac{\partial }{\partial r} - \mu_0 \nu_m\sigma\right)E_z = \mu_0\left( \frac{\partial J_b}{\partial t} - \nu_m J_p\right).
\end{eqnarray}
The electric field and current densities are discretized onto a radial grid that extends from 0 to 8.65~cm with 100 grid points.
The operator acting on $E_z$ on the left-hand side of this equation is written as a finite difference matrix, and the electric field is updated at each time step using the \texttt{solve} function from the SciPy package \texttt{scipy.linalg}.\cite{2020SciPy-NMeth} This function uses the industry-standard \texttt{LAPACK} library to solve the matrix equation using LU decomposition. The values of the electric field and plasma current at time step $n$ are used when calculating the plasma current at time step $n+1$:
\begin{eqnarray}
J_p^{n+1} = J_p^{n} + \Delta t \nu_m \left( \sigma E_z^{n} - J_p^{n}\right).
\end{eqnarray}

The rate equations for the density of the electrons, ions, ground state neutral N$_2$, and all excited states are solved numerically, using reaction rates that are tabulated as functions of the mean electron energy. The mean electron energy is obtained by solving Eq.~\eqref{electron_energy_eqn} numerically. The forward Euler method is used for solving these density and energy equations.

The simulations shown in the remainder of this section were produced using various open-source packages, most notably turboPy, NumPy, Matplotlib, and xarray.\cite{turbopy_code,turbopy_code_v2020.10.14,harris2020array,Hunter:2007,hoyer2017xarray}

\subsection{Simulation results}

To compare the results from the rigid-beam model with experiments, the rigid-beam turboPy program described above was run with background gas pressures of 0.7, 1, 3, 5, 7, and 10~Torr. 
The grid spacing and size of the time step were chosen so that the simulation is numerically converged (see the Appendix for details). While the experiments were performed in dry air, and the model only includes nitrogen, the model results are expected to at least qualitatively reproduce the experimental measurements.

\begin{figure}
\includegraphics[width=\columnwidth]{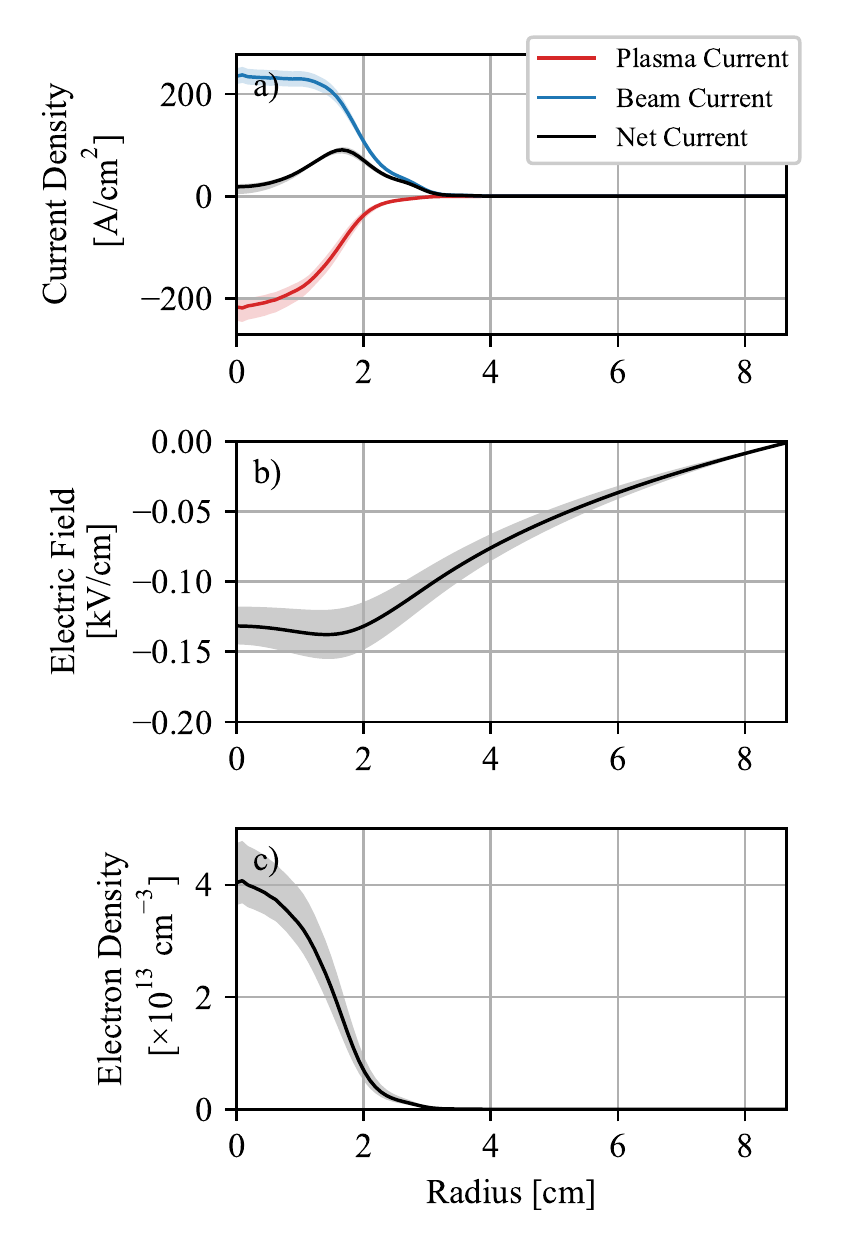}
\caption{
\label{profile}Radial profiles at 40~ns of a) the simulated axial current densities, b) the axial electric field, and c) the electron density. Results are from the simulation with 1~Torr background gas pressure. 
}
\end{figure}

Figure~\ref{profile} shows radial profiles of several quantities during the beam rise, at 40~ns, from the simulation with 1~Torr background gas pressure. The beam current (shown in blue in Fig.~\ref{profile}a) is confined within about 3~cm radius, as are the plasma electrons (Fig.~\ref{profile}c). The axial electric field (Fig.~\ref{profile}b) drives an axial current in the plasma (shown in red in Fig.~\ref{profile}a), which cancels some of the beam current, resulting in a net axial current  (shown in black in Fig.~\ref{profile}a) which peaks near a radius of 1.7~cm.

\begin{figure}
\includegraphics[width=\columnwidth]{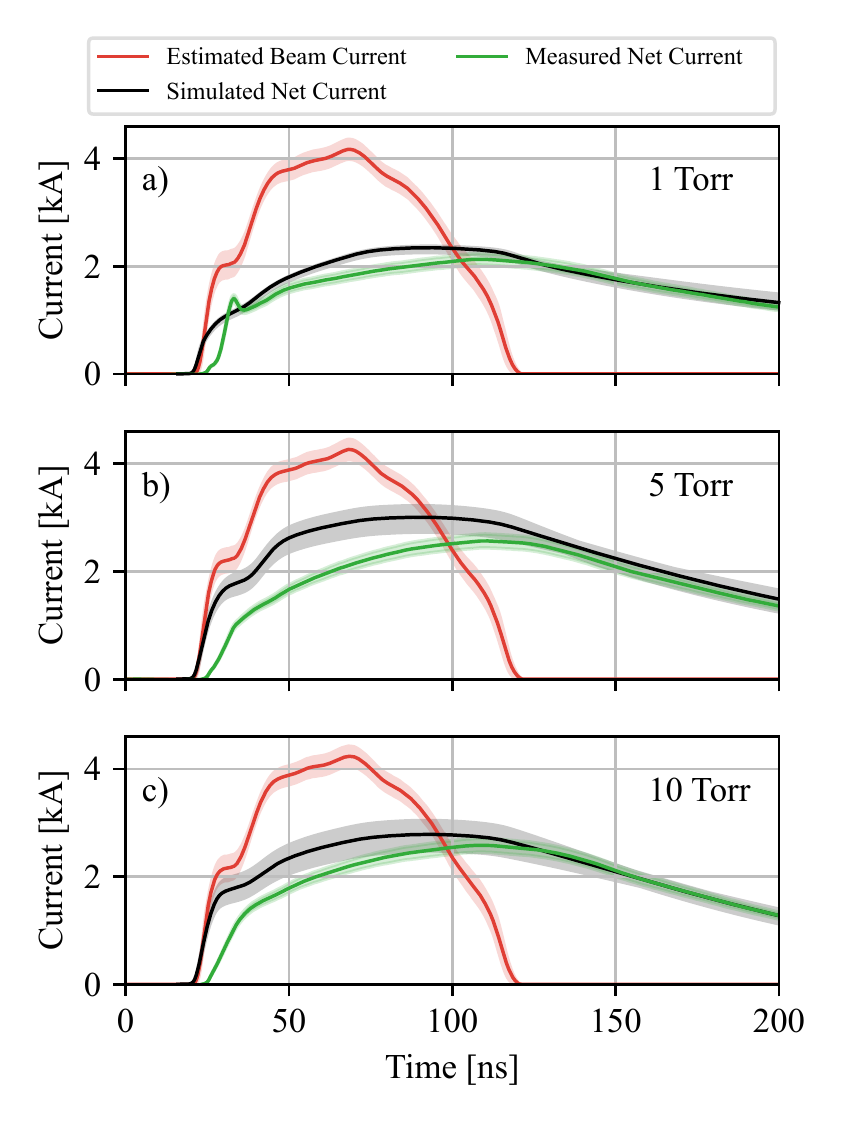}
\caption{\label{currents}Comparison of estimated beam current (red), measured net current (green), and simulated net current (black) for a) 1~Torr, b) 5~Torr, and c) 10~Torr. 
}
\end{figure}

Figure~\ref{currents} shows a comparison of the measured net current\footnote{Data at 1, 5, 10~Torr from Shots 1651, 1655, 1657, respectively.} (green), the beam current that drives this simulation (red), and the simulated net current (black), at 1, 5, and 10 Torr. The net current initially rises with the beam current, but then the plasma current starts to cancel out a portion of the beam current, giving a lower net current. Once the beam pulse has ended, the inductive electric field keeps the current flowing in the plasma and the plasma current slowly decays away because of the finite plasma resistivity. The differences between the simulated and measured net currents, especially at early times, could be due to approximations in the model or in the current pulse used to drive the simulations. These differences will be examined more closely in future work.

\begin{figure}
\includegraphics[width=\columnwidth]{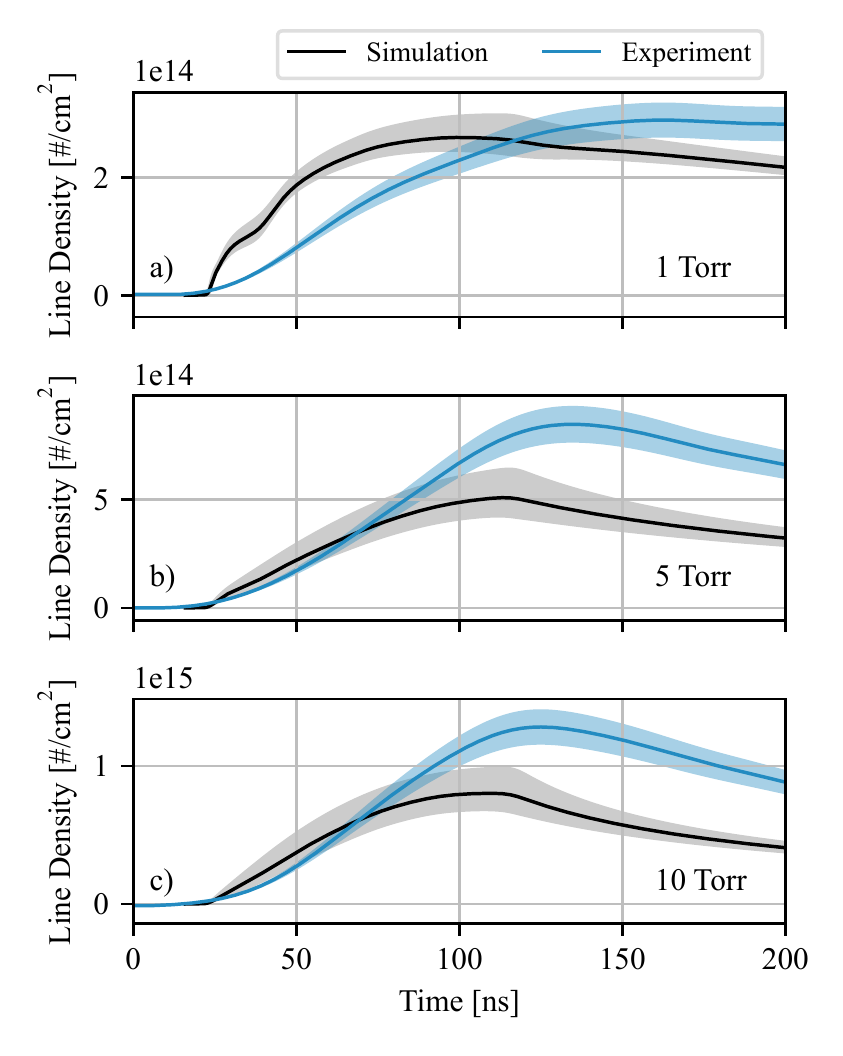}
\caption{\label{density}Comparison of measured (blue) and simulated (black) on-axis line-integrated electron density for a) 1 Torr, b) 5 Torr, and c) 10 Torr.
}
\end{figure}

The line-integrated electron density is measured in the experiments using an interferometer with a line-of-sight along a chord through the plasma. A synthetic diagnostic was used in the simulation to obtain a similar output, where the chord is taken through the center of the plasma. Figure~\ref{density} shows this line-integrated plasma electron density as a function of time, for both simulation (black) and experiment (blue) at 1, 5, and 10~Torr. The generation of plasma through beam-impact and thermal ionization gives densities that rise during the beam pulse, and then dissociative recombination causes the densities to decrease later in time. At lower gas pressures, beam-impact ionization has a smaller effect, and the density peaks at a smaller value. These trends are seen in both the experimental data and the simulation outputs. The simulations  reproduce the experimental trends with gas pressure, and the simulated peak value of line-integrated density is about a factor of two lower than the experiment. There are several factors that could contribute to these differences, such as uncertainties in the rates used in the plasma model, or the fact that the model only includes nitrogen but the experiments were in air. Further work is planned to study these differences.

\begin{figure}
\includegraphics[width=\columnwidth]{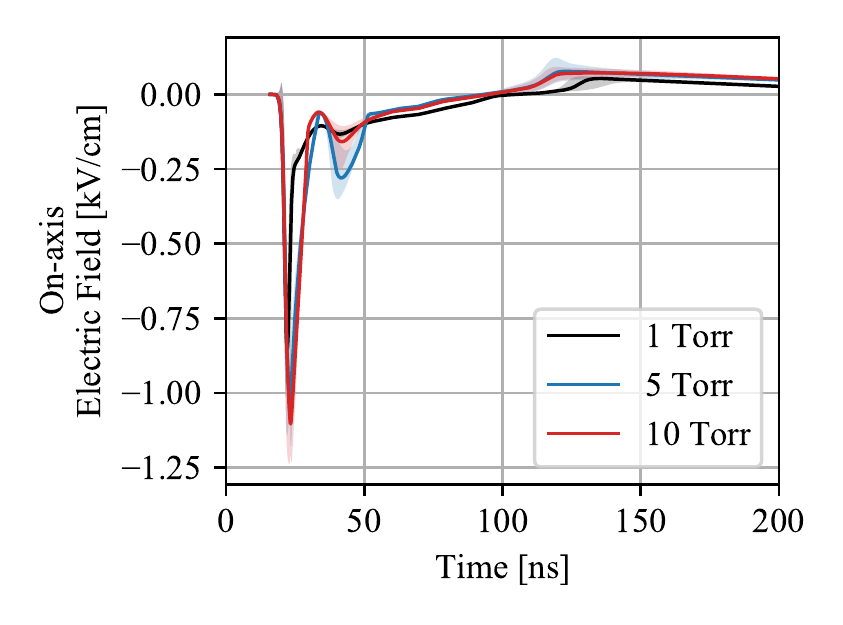}
\caption{\label{efield}On-axis electric field from 1, 5, and 10~Torr simulations.}
\end{figure}

In addition to experimentally motivated synthetic diagnostics, additional diagnostics can be  added to the simulation in order to probe the dynamics of the model. For example, the value of the induced electric field on axis as a function time is shown in Fig.~\ref{efield} for the 1, 5, and 10 Torr simulations. The rapidly rising beam current density initially induces a very large electric field. As the plasma becomes more conductive, the plasma current tends to respond quickly enough to cancel out rapid changes in the beam current, and thus the magnitude of the induced electric field is smaller later in time. Once the beam pulse has ended, the stored magnetic energy in the chamber slowly dissipates through the resistive plasma. The electric field associated with this dissipating energy can be seen in the long, late time tail in Fig.~\ref{efield}, and is generally much smaller than the peak field early in the pulse. 

\begin{figure}
\includegraphics[width=\columnwidth]{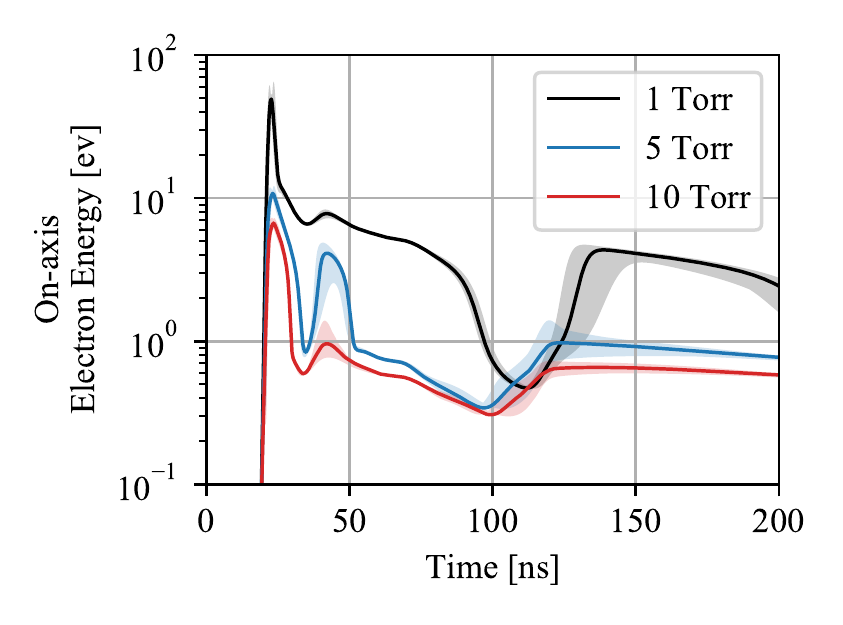}
\caption{\label{energy}Mean electron energy $U_e$ on axis from the 1, 5, and 10~Torr simulations.}
\end{figure}

The plasma response model can also be probed through the use of custom diagnostics. Figure~\ref{energy} shows the mean energy of the plasma electrons on axis for the 1, 5, and 10~Torr simulations, with peaks in energy that correlate to the induced electric fields, showing Ohmic heating of the plasma electrons. The peak at early time, around 20~ns, corresponds to the large electric field that is induced during the rise of the beam current. The smaller peak at late time, around 120~ns, corresponds to the end of the beam pulse, which also induces an electric field which heats the electrons.

\begin{figure}
\includegraphics[width=\columnwidth]{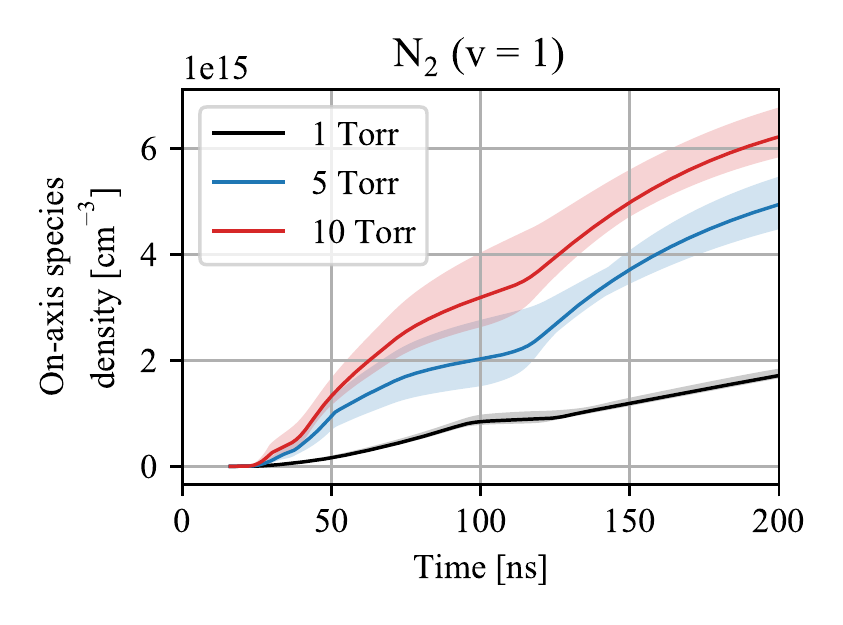}
\caption{\label{species_v1}On-axis density of the vibrationally excited state N$_2$($v=1$), for the 1, 5, and 10 Torr simulations
}
\end{figure}

\begin{figure}
\includegraphics[width=\columnwidth]{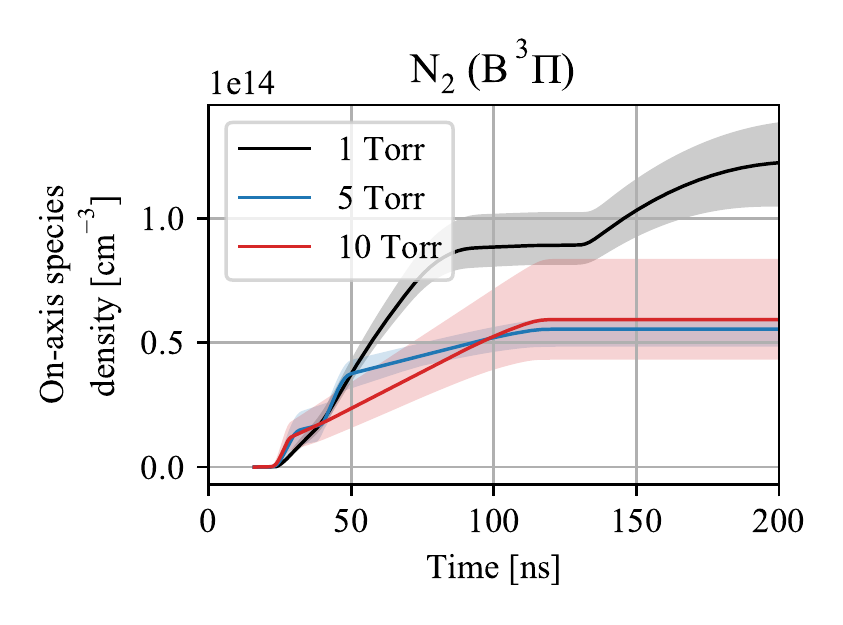}
\caption{\label{species_B3}On-axis density of the electronically excited state N$_2$(B$\,^3\Pi$), for the 1, 5, and 10 Torr simulations}
\end{figure}

In Figs.~\ref{species_v1} and \ref{species_B3}, the time histories of the on-axis densities of the N$_2$($v=1$) vibrational state and the N$_2$(B$\,^3\Pi$) electronic state of nitrogen are shown for simulations at 1, 5, and 10~Torr. These states exhibit changes in density due to both beam-impact and thermal excitation. As can be seen by the increasing densities late in time, the thermal plasma electrons continue to excite these states even after the pulse ends. At 10~Torr, excitation of the N$_2$(B$\,^3\Pi$) happens almost exclusively through beam-impact events, which can be understood by examining the energy of the thermal electrons in Fig.~\ref{energy}. This state has a threshold energy of 7.35~eV, but the electrons in the 10~Torr simulation are significantly lower energy than this for nearly the entire simulation.
Note that the model used for these simulations does not include any means by which the densities of the excited neutral states could decrease. Future work should include, for example, the physics of collisional de-excitation in order to more accurately track the densities of these states and estimate optical emission spectra.
The densities of all the excited states at 125~ns (near the peak of the line-integrated densities) are summarized in Table \ref{tab_final_density} for the 10 Torr simulation.
\renewcommand{\arraystretch}{1.1}
\begin{table}
\caption{\label{tab_final_density}Simulated density from 10~Torr simulation.}
\begin{ruledtabular}
\begin{tabular}{lcccc}
Species & Density at 125~ns [\%]\footnotemark[1]\\
\hline
X$\,^1\Sigma$ & 95.15 \\[2pt]
{\em rot} & 2.51\\[4pt]
$v=1$, 2, 3, 4 
	&  1.12, 0.36, 0.24, 0.14 \\
$v=5$, 6, 7, 8 
	&  0.11, 0.08, 0.04, 0.02\\[4pt]
A$\,^3\Sigma$($v={0-4}$), A$\,^3\Sigma$($v={5-9}$) 
& 0.01, 0.01 \\
B$\,^3\Pi$, W$\,^3\Delta$, A$\,^3\Sigma$($v\ge{10}$), B$^\prime\,^3\Sigma$ 
& 0.02, 0.01, 0.01, 0.00 \\
a$^\prime\,^1\Sigma$, a$\,^1\Pi$, w$\,^1\Delta$, C$\,^3\Pi$ 
& 0.00, 0.01, 0.00, 0.01
\\
E$\,^3\Sigma$, a$^{\prime\prime}\,^1\Sigma$, {\em sum} 
& 0.00, 0.00, 0.07
\\[4pt]
 N$_2^+$ & 0.05\\[4pt]
Atomic N & 0.10
\end{tabular}
\footnotetext[1]{Density is expressed as a percent of the initial neutral density, $3.53\times10^{17}\text{ cm}^{-3}$.}
\end{ruledtabular}
\end{table}

\subsection{Simulation results as functions of pressure}\label{sims_vs_pressure}
\begin{figure}
\includegraphics[width=\columnwidth]{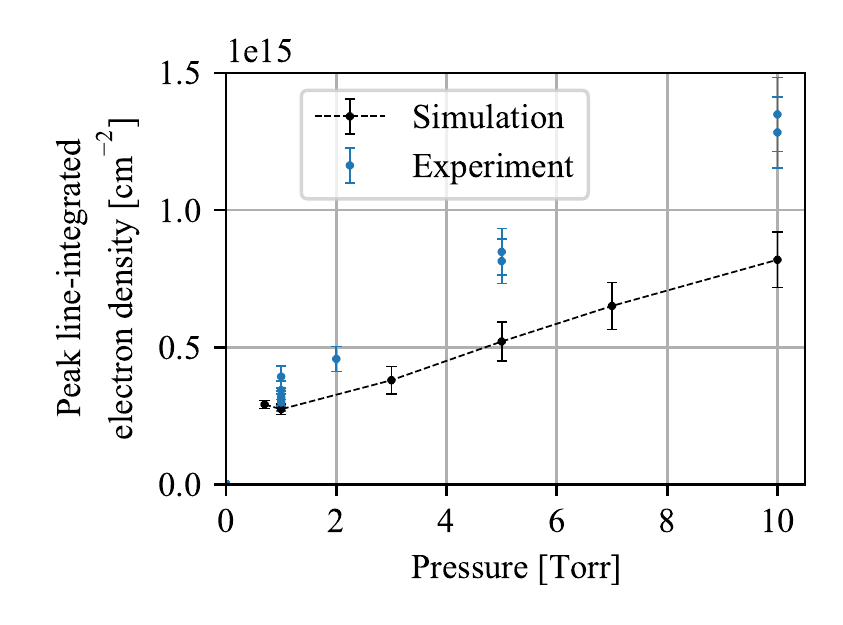}
\caption{\label{peak_density_vs_pressure}Peak value of line-integrated electron density vs pressure. Experimental measurements include $\pm 10$\% uncertainty bars, and data from multiple shots at the same pressure are plotted separately to show experimental reproducibility. 
}
\end{figure}

\begin{figure}
\includegraphics[width=\columnwidth]{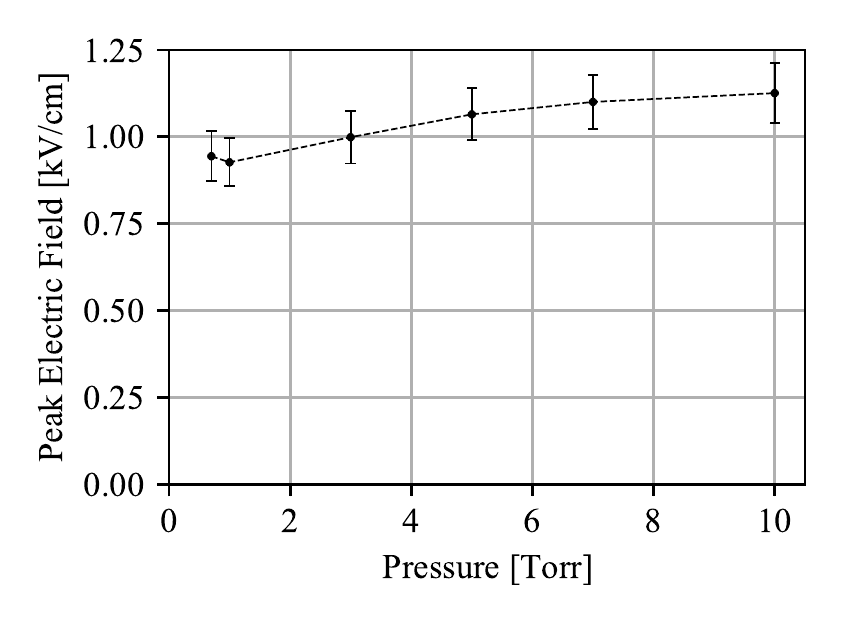}
\caption{\label{peak_electric_field}Peak value of the on-axis electric field vs pressure.
}
\end{figure}

\begin{figure}
\includegraphics[width=\columnwidth]{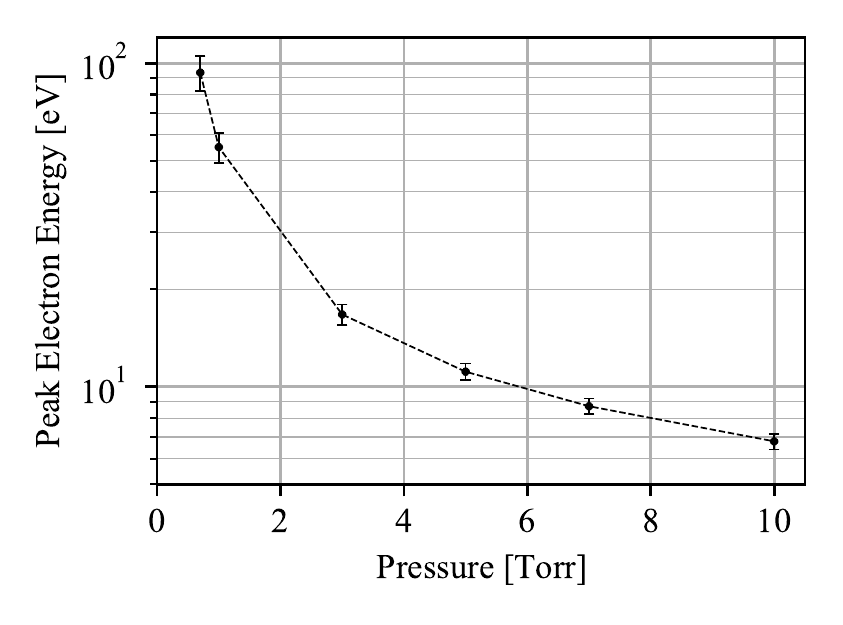}
\caption{\label{peak_electron_energy}Peak value of the on-axis electron mean energy $U_e$ vs pressure.
}
\end{figure}

\begin{figure}
\includegraphics[width=\columnwidth]{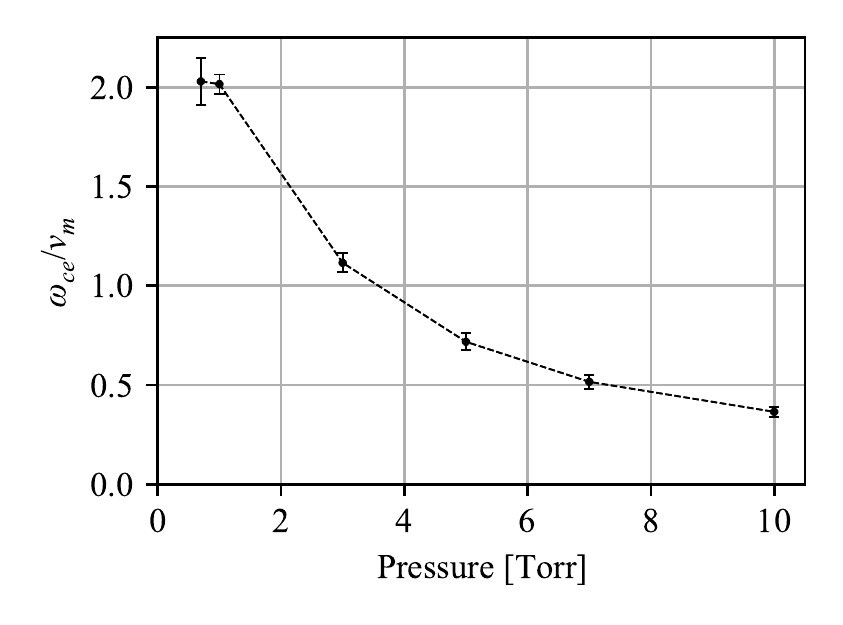}
\caption{\label{peak_hall}Peak value of the Hall coefficient C$_\text{H} = \omega_\text{ce}/\nu_m$, as a function of pressure.
}
\end{figure}

\begin{figure}
\includegraphics[width=\columnwidth]{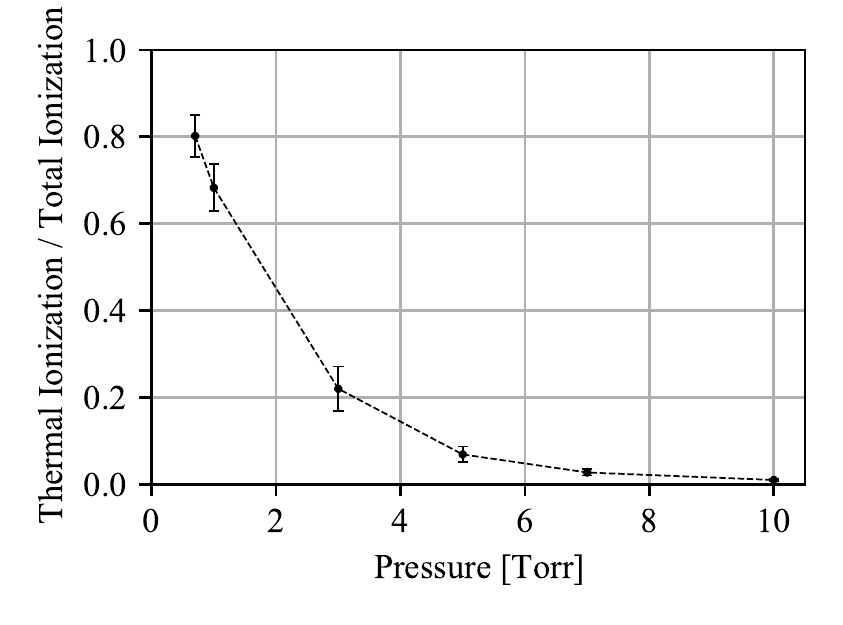}
\caption{\label{thermal_fraction}Fraction of on-axis ionizations due to thermal processes vs pressure at 200~ns. At pressures above about 2~Torr, beam-impact dominates the plasma formation.
}
\end{figure}

\begin{figure}
\includegraphics[width=\columnwidth]{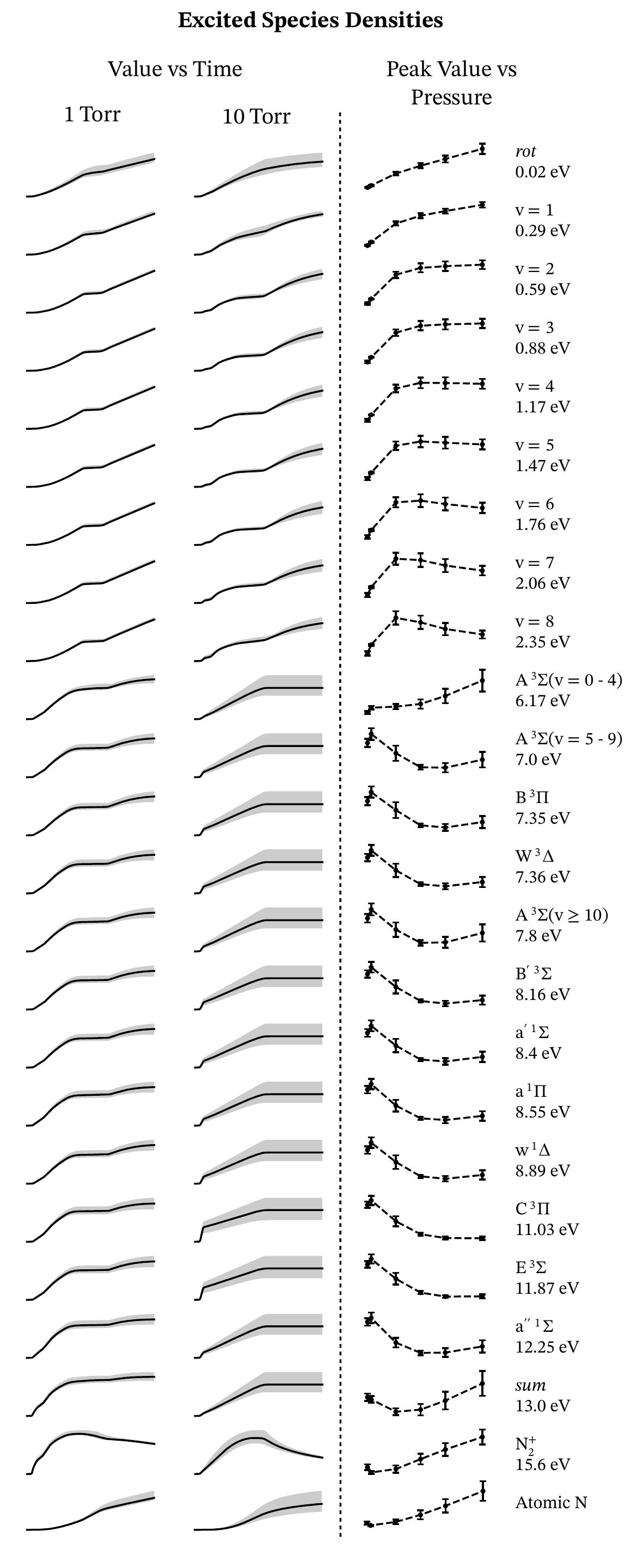}
\caption{\label{fig_sparklines}
Sparklines showing the trends of (first column) density vs time for 0--200~ns for $p$ = 1~Torr, (second column) density vs time for $p$ = 10~Torr, and (third column) maximum density vs pressure for 0.7--10~Torr. Rows are sorted in order of increasing threshold energy. Note: beam-impact excitation of the A$\,^3\Sigma$ state is distributed evenly between the three vibrational bands.
}
\end{figure}

Because of the simplifications and reductions in the rigid-beam model, the numerical simulations described above run quickly, and parameter sweeps can be performed without significant computational cost. This allows for the rapid evaluation of chemistry models, and pressure scans can be easily performed. To illustrate this, a set of simulations were performed for gas pressures ranging from 0.7~Torr to 10~Torr. For each simulation, the peak values of several quantities are recorded, and are shown as functions of pressure in Figs.~\ref{peak_density_vs_pressure} -- \ref{fig_sparklines}.  At each pressure, the dots show the mean of the nine simulations with different combinations of beam current and energy histories, and the error bars show standard deviation. The dashed lines connect the mean values to illustrate the variation with pressure. Note that simulations at pressures lower than 0.7~Torr are not shown, because at those low pressures the mean plasma electron energy reaches values greater than 100~eV. This is the maximum energy value for the reaction rates tables used in these simulations, and reaching this value indicates that the chemistry model used in this paper is likely insufficient at these low pressures.

Figure~\ref{peak_density_vs_pressure} shows the peak value of the line-integrated electron density. Since the beam-impact ionization rate is proportional to the number density of the background gas, the density of plasma produced by beam-impact ionization should increase with pressure. As seen in the figure, this is what happens for pressures above about 3~Torr. At lower pressures, the effect of thermal ionization starts to become more important, and the density is higher than would be expected from beam-impact ionization alone. The trend of density increasing with gas pressure in the experiments is captured in the simulations. However, the peak line-density from the simulations is lower than the measured value at all pressures. This discrepancy is slightly larger than the $\pm 10$\% uncertainty in the measurements. 
The overall comparison of the model with the data is within a factor of about 1.5 at all pressures and the experimental trends are reproduced. This is an indication that the model captures the essence of the physics and provides an excellent framework to do systematic studies of the beam-plasma system. An effort is underway to further investigate the source of the remaining discrepancy. Also, note that the experimental data at pressures lower than 0.7~Torr are outside the range of model validity, and are thus not shown in Fig.~\ref{peak_density_vs_pressure}.

Figure \ref{peak_electric_field} shows the peak value of the on-axis electric field. While this value does vary somewhat with pressure, the values do not vary by more than about 20\%. This is not the case for the mean electron energy, however. Figure \ref{peak_electron_energy} shows the on-axis value of the mean electron energy. This quantity does strongly depend on pressure, varying by nearly a factor of 10, from a minimum of about 6~eV at 10~Torr to a maximum of above 100~eV at 0.7~Torr. For the simplified plasma chemistry that is being used in this work, there are important electron scattering processes at higher energy that have been neglected. At mean energies above about 10~eV, there are likely to be processes such as stepwise ionization or ionization to excited states of N$_2^+$ which can significantly affect the electron energy. The results in Fig.~\ref{peak_electron_energy} indicate that this simplified chemistry should be used with care below about 5~Torr, and below about 2~Torr the model could be giving incorrect results because of the missing plasma chemistry. The results at 0.7~Torr give peak energies above 100~eV. Since the rate tables used in the work only extend to 100~eV, these results indicate that the model should not be used at any lower pressures.

The magnetic field in each simulation was calculated from the net current using Amp\`ere's Law  Eq.~\eqref{ampere}. This was used to calculate the electron cyclotron frequency so that the Hall coefficient C$_\text{H}$ for each simulation could be calculated.  As shown in Fig.~\ref{peak_hall}, the maximum value (for any radius and time) of C$_\text{H}$ is less than one at higher pressures but above one for simulations with pressures of 3~Torr and below. This is an indication that the assumptions made in this model may not be valid at these lower pressures, and that the $\V_e\times\B$ term in the momentum equation should be retained.

There is another quantity that can be extracted from the simulations in order to obtain insights into the physics of formation of the plasma. Since thermally driven ionizations are computed separately from beam-impact ionizations, the code can track how much of the plasma was created by thermal processes and how much by beam impact. The fraction of ionization events due to thermal processes is calculated by taking the ratio of electron density produced thermally to the total electron density. The on-axis value of this ratio at 200~ns is show in Fig.~\ref{thermal_fraction} as a function of pressure. Above about 2~Torr, the fraction of plasma produced thermally drops below 50\%, indicating that at higher pressure beam-impact ionization is the dominant process by which the plasma is produced. This is consistent with the approximately linear scaling of peak plasma density with pressure shown in Fig.~\ref{peak_density_vs_pressure}.

The last simulation output that is useful to examine is a summary of the dynamics of the excited states of N$_2$. Plots such as those in Figs.~\ref{species_v1} and \ref{species_B3} can be helpful for predicting experimental outputs that depend on the densities of excited states, such as spectral measurements of the optical emission once the necessary processes are included. In order to condense this information in a way that would be helpful for discerning trends, a series of sparkline plots were created and are shown in Fig.~\ref{fig_sparklines}. Each row is a different state, and the states are sorted in order of increasing threshold energy (the state name and threshold energy are listed to the right of each corresponding row). The left column of this plot shows the density vs time sparkline from the 1~Torr pressure simulation. The second column shows density vs time from the 10~Torr simulation. This column shows some differences from the 1~Torr case. Most noticeably the fact that beam-impact excitation is more important for the states with threshold above $\sim$6~eV at 10~Torr than it is for 1~Torr. This is seen in the difference in the shapes of the density curves; in the 1~Torr case the density of these states continues to increase after the pulse is over, which for the 10~Torr case the density is constant after the pulse.  The third column of sparklines shows how the peak value of density varies with pressure. These are sparkline versions of plots that are analogous to the plot in Fig.~\ref{peak_density_vs_pressure} for electron density. Several of the species, such as the N$_2^+$ ion and the N$_2$({\em rot}) state, increase with pressure. Others, however, such as the B$\,^3\Pi$ and C$\,^3\Pi$ electronic states, actually decrease with pressure. It would be interesting if these trends could be observed experimentally through spectroscopic or other means. In order to use this rigid-beam code to predict optical spectra, a more complete model of the plasma chemistry is required, including a model of the radiation dynamics.

Plots like Fig.~\ref{fig_sparklines} can be valuable for identifying areas where the plasma chemistry model should be improved. For example, in the $p=10$~Torr case, densities for all excited states with thresholds above 6.17~eV are unphysically constant after the beam is off, showing clearly the impact of neglecting de-excitation processes. 

\section{Conclusions}\label{sec_conclusions}

In this paper, we define a reduced model of the electrical breakdown of a low-pressure gas caused by the injection of an intense electron beam. Several simplifying approximations are made to the equations for the beam dynamics and the electromagnetic fields. Much of the physics is driven by the electric field that is induced by the changing beam current, and the {\em rigid-beam model} expresses the dynamics of this field through a 1-dimensional second-order Poisson-like differential equation. 

The rigid-beam model, when combined with a model for plasma that is created by beam-impact reactions and reactions driven by the electric field, forms a complete set of equations for the dynamics of the beam/plasma system. A fluid model of the plasma electrons is presented, along with a weakly ionized plasma chemistry which describes the electron-driven scattering processes in low pressure molecular nitrogen. The prescribed properties of the injected beam, provided by experimental measurements, appear in source terms to the electron fluid equations.

The coupled rigid-beam equations and the plasma response are solved with a code written using the turboPy framework. Numerical results are presented for experimentally relevant parameters, showing that this model is useful for computing the beam-driven breakdown of a low-pressure gas. Because of the simplifications inherent to the rigid-beam model, the code runs quickly. This enables rapid numerical exploration of the parameter space, which includes variation with pressure, electron beam current density, and the plasma chemistry model. In addition to running simulations for a range of model parameters, it is also straightforward to expand the code through additional PhysicsModules which implement different reduced plasma models. 

There are several ways in which the model described in this paper expands on previous work. Previous papers which used a rigid-beam model to simulate e-beam driven breakdown had several limitations which prevented them from accurately modeling the experiments described in this paper. Ref.~\onlinecite{doi:10.1063/1.4950840} used a 0-D model for the beam, which has been expanded to a 1-D radial model in this paper. Ref.~\onlinecite{swanekamp2021rigidbeam} used a simplified plasma chemistry, which has been extended here to include beam-impact ionization and dissociative recombination. Additionally, a more realistic radial profile and temporal pulse shape have been used for the beam. By extending the rigid-beam model in these ways, higher-fidelity simulations are able to be perfomed.
Even with all the simplifications in this model, the simulation results agree favorably with experimental measurements. The simulated line-integrated electron density increases with increasing pressure, which is a trend observed in experiments. While the simulated densities are consistently lower than measured in experiments, the results are within a factor of two of the measurements.

Future work is planned where this program will be used to map out regions of validity in parameter space for the reduced plasma response models used in this work. Extensions to the program are also planned, which will implement models that account for physics which has been neglected. For example, electron-ion collisions will be added to the conductivity, de-excitation processes will be added for tracking excited state populations, and advanced chemistries will be added which include state-to-state transitions.
 Additionally, this program can be used to perform sensitivity studies, for example to determine the effect of uncertainties in the electron-scattering cross sections that are used in the weakly ionized plasma chemistry.

The RB model described in this paper employs a set of assumptions about the electromagnetic fields and the electron-beam dynamics which are useful for modeling the breakdown of low-pressure gas by an intense electron beam. This model also lays a foundation for future work studying approximations related to the plasma electron dynamics or related to the plasma chemistry. Systematic investigation of advanced plasma chemistries---which examine the effects of state-to-state transitions for example---can now be performed and compared {\em ceteris paribus} to results obtained with simpler plasma chemistry approximations.

\begin{acknowledgments}
The authors wish to thank Dr.~P.~Ottinger and Dr.~T.~Haines for useful discussions.

This work was funded by the Defense Threat Reduction Agency and the US Naval Research Laboratory Base Program.
\end{acknowledgments}

\section*{Data Availability}
The data that support the findings of this study are available from the corresponding author 
upon reasonable request.

\appendix*

\section{Convergence}
\label{convergence}

The grid and the time step are tested at various pressures in order to ensure that the numerical solution is acceptably converged. The asymptotic approach in Ref.~\onlinecite{10.1115/1.2910291} allows us to relate the approximate result to the expected result as we refine the grids in space and time. The method is a uniform refinement method generalized from Richardson extrapolation theory. Here, the grid convergence index (GCI) gives some statistical significance to the percent error away from the converged solution. The advantages of using this method for the rigid-beam problem are that it allows for quick and confident testing of space and time resolution used to produce the results in the paper.

The electric field on axis was examined in this grid spacing convergence study. The electric field in the rigid-beam model is computed using a second order discretization in space, thus this is a natural quantity to use for this convergence study. {The grid extends from $r=0$ to $r=8.65$~cm.} The grid spacing is refined from 10 grid points, to 30 grid points, to 90 grid points and the results at different pressures are reported in Table \ref{convergence1}.
\begin{table}
\caption{\label{convergence1} %
Convergence with respect to grid spacing
}
\begin{ruledtabular}
\begin{tabular}{cccc}
\thead{Pressure} & \thead{Order of\\Convergence ($\mathcal{O}$)} 
& \thead{Extrapolated\\Value $h_0$} & \thead{GCI \%} \\ 
\hline
1 & 1.677 & 475 & 8.71 \\
5	&1.579	&2063	&1.57 \\
7	&1.675	&2159	&0.857\\
10	&1.704	&1991	&0.614\\
\end{tabular}
\end{ruledtabular}
\end{table}
Here, the theoretical order of convergence is $\mathcal{O}(2)$ and what we found as the actual order of convergence is $\mathcal{O}(1.6)\sim\mathcal{O}(1.7)$. The extrapolated value $h_0$ is what the approximate converged value of the electric field is on axis at the end of the simulation, and the GCI is the error band from the fine grid (90 points) to the medium grid (30 points). The results show that as the pressure increases the error gap improves and the solution is closer to the converged value, but all pressures are acceptably converged.

Because of the optimized solver, vectorized operations, and relatively small number of grid points used in these simulations, increasing the number of spatial grid points did not affect the runtime of the simulations very much. For the 1~Torr simulation with time step of $1\times 10^{-3}$~ns, the simulations with 10, 30, and 90 grid points took 441.6, 442.9, and 438.9 seconds to run, respectively.

Similarly, for time step refinement, we choose the net current as the variable to study for this experiment. In this experiment, we refine the time step from $2\times10^{-3}$~ns to $1\times10^{-3}$~ns, and to $5\times10^{-4}$~ns. Here, we expect the theoretical order of converge to be $\mathcal{O}(1)$ because of the forward Euler method used for the time update. The results at different pressures are reported in Table \ref{convergence2}.
\begin{table}
\caption{\label{convergence2} %
Convergence with respect to time step size
}
\begin{ruledtabular}
\begin{tabular}{cccc}
\thead{Pressure} & \thead{Order of\\Convergence ($\mathcal{O}$)} 
& \thead{Extrapolated\\Value $h_0$} & \thead{GCI \%} \\
\hline
1	&0.9424	&2354	&0.42\\
5	&1.008	&3313	&0.111\\
7	&0.9855	&3307	&0.208\\
10	&0.9824	&3240	&0.351\\
\end{tabular}
\end{ruledtabular}
\end{table}
The results show the actual order of convergence is almost exactly the predicted value and the error bands of this experiment are all under 1\% with the best case being the 5 Torr result. 

Changing the size of the time step used in these simulations had a much larger affect on the run time. Simulations with 100 grid points at 1~Torr pressure and time steps of $2\times10^{-3}$~ns, $1\times10^{-3}$~ns, and $5\times10^{-4}$~ns took 240.7, 485.6, and 949.4 seconds to run, respectively.

For the rigid-beam model and the results shown in this paper, we find that the use of 90 grid points and time step of $5\times10^{-4}$~ns is acceptable. For the results shown above in Section \ref{sec_results}, we used  $5\times10^{-4}$~ns as the time step and rounded up to 100 grid points for the spatial resolution.

\nocite{*}
\bibliography{rb}

\end{document}